\documentclass[11pt,a4paper]{emulateapj}

\usepackage{psfrag}
\usepackage{psfig}
\usepackage{pspicture}
\usepackage{graphicx}

\usepackage{amssymb,amsmath}
\usepackage{natbib}

\def\apj{ApJ}

\journalinfo{\@Accepted for publication in the \apj}
\submitted{Accepted for publication in the \apj}

\begin{document}

\title {A Comparative Study of Density Field Estimation for Galaxies: New Insights into the Evolution of Galaxies with Environment in COSMOS out to \lowercase{$z\sim$}3}

\author{
Behnam Darvish,\altaffilmark{1}
Bahram Mobasher,\altaffilmark{1}
David Sobral,\altaffilmark{2,3,4}
Nicholas Scoville,\altaffilmark{5}
Miguel Aragon-Calvo\altaffilmark{1}
}

\setcounter{footnote}{0}
\altaffiltext{1}{University of California, Riverside, 900 University Ave, Riverside, CA, 92521, USA; email: bdarv001@ucr.edu}
\altaffiltext{2}{Instituto de Astrof\'{\i}sica e Ci\^encias do Espa\c co, Universidade de Lisboa, OAL, Tapada da Ajuda, PT 1349-018 Lisboa, Portugal}
\altaffiltext{3}{Centro de Astronomia e  Astrof\'{\i}sica da Universidade de Lisboa, Observat\'{o}rio Astron\'{o}mico de Lisboa, Tapada da Ajuda, 1349-018 Lisboa, Portugal}
\altaffiltext{4}{Leiden Observatory, Leiden University, P.O. Box 9513, NL-2300 RA Leiden, The Netherlands}
\altaffiltext{5}{California Institute of Technology, MC 249-17, 1200 East California Boulevard, Pasadena, CA 91125, USA}

\begin{abstract}

It is well-known that galaxy environment has a fundamental effect in shaping its properties. We study the environmental effects on galaxy evolution, with an emphasis on the environment defined as the local number density of galaxies. The density field is estimated with different estimators (weighted adaptive kernel smoothing, 10$^{th}$ and 5$^{th}$ nearest neighbors, Voronoi and Delaunay tessellation) for a K$_{s}<$24 sample of $\sim$190,000 galaxies in the COSMOS field at 0.1$<$z$<$3.1. The performance of each estimator is evaluated with extensive simulations. We show that overall, there is a good agreement between the estimated density fields using different methods over $\sim$2 dex in overdensity values. However, our simulations show that adaptive kernel and Voronoi tessellation outperform other methods. Using the Voronoi tessellation method, we assign surface densities to a mass complete sample of quiescent and star-forming galaxies out to z$\sim$3. We show that at a fixed stellar mass, the median color of quiescent galaxies does not depend on their host environment out to z$\sim$3. We find that the number and stellar mass density of massive ($>$10$^{11}$M$_{\odot}$) star-forming galaxies have not significantly changed since z$\sim$3, regardless of their environment. However, for massive quiescent systems at lower redshifts (z$\lesssim$1.3), we find a significant evolution in the number and stellar mass densities in denser environments compared to lower density regions. Our results suggest that the relation between stellar mass and local density is more fundamental than the color-density relation and that environment plays a significant role in quenching star formation activity in galaxies at z$\lesssim$1.

\end{abstract}

\keywords{}

\section{Introduction} \label{intro}

It is known that the environment in which galaxies reside plays a fundamental role in their evolution. In the local universe, denser environments are dominated by red, passive, early-type galaxies whereas less dense regions are preferentially populated by blue, star-forming, late-type systems \citep{Dressler80,Kauffmann04,Balogh04,Baldry06,Bamford09,Peng10}. These environmental trends still hold at higher redshifts \citep{Capak07b,Cooper07,Peng10,Sobral11,Muzzin12,Scoville13}, although they usually tend to weaken with increasing redshift. While it is evident that almost any observable property of a galaxy demonstrates some association with environment, there is still a question that needs to be addressed first. What do we really mean by ``environment"?\\
Recent advances in numerical simulations (Millennium, \citealp{Springel05}; Illustris, \citealp{Vogelsberger14}) combined with extensive spectroscopic observations of local galaxies (SDSS, \citealp{York00}; 2dFGRS, \citealp{Colless01}; GAMA, \citealp{Driver11}) have revealed that the universe has a web-like pattern, i.e.; the ``cosmic web" \citep{Bond96}, containing dark matter, gas and luminous galaxies. Galaxies in the cosmic web are organized in a network containing dense clusters, sparsely populated voids, planar walls, and thread-like filamentary structures linking overdense regions. Therefore, the most natural approach in defining the environment of a galaxy is to locate it within the cosmic web. However, the complexity and lack of a fully objective method in identifying the major components of the cosmic web often limit the environmental studies of galaxies within the comic web to numerical simulations or large spectroscopic surveys in the local universe.\\ 
The conventional method of defining the environment as two extreme regions in the density distribution of galaxies, i.e. galaxy cluster and the general field, does not usually account for the full dynamical range of the density field. For example, there are intermediate environments such as galaxy groups, outskirts of clusters and filaments which are equally important \citep{Fadda08,Porter08,Tran09,Geach11,Coppin12,Mahajan12,Pintos-Castro13,Darvish14}. Moreover, the selection of galaxies to whether they belong to the cluster environment or the general field is somewhat subjective.\\
The detection of galaxy clusters through their diffuse X-ray emission \citep{Gioia90,Ebeling98,Bohringer00,Ebeling01,Bohringer04,Mehrtens12}, the red sequence and its variants \citep{Gladders00,Goto02,Gladders05,Miller05,Koester07,Hao10,Muzzin13b}, weak gravitational lensing \citep{Miyazaki07,Gavazzi07,Dietrich07} and the Sunyaev-Zel'dovich effect \citep{Planck11,Williamson11,Marriage11,Reichardt13} adds more assumptions and limitations in the environmental studies. For example, weak lensing, hot X-ray emission and the Sunyaev-Zel'dovich effects are more sensitive to virialized, massive galaxy clusters. The red sequence technique, which relies on observations in only two filters, is very successful and economically efficient in detecting a large sample of galaxy clusters that have a tight red sequence of quiescent galaxies. However, this technique assumes the existence of a tight red sequence for clusters, requires modelling the red sequence and uses only the quiescent systems as a proxy of their host cluster environment. Moreover, the quiescent galaxies become less abundant at higher redshifts which makes the cluster detection techniques based on the red sequence (or galaxy color) challenging at high redshift. All these methods may produce biased environments, which may lead to misinterpretation of the evolutionary history of galaxies as a function of their host environment.\\
Another approach in defining environment is to use the local number density (usually surface density) of galaxies as a proxy to their host region. Among the most common density estimators used in the literature are the nearest neighbor (usually 5$^{th}$ or 10$^{th}$  NN) and count-in-cell (CC) methods. However, although simple to implement and computationally fast, the performance of these methods strongly depends on the depth of the data, the number of neighbors considered in the analysis (in NN method) and the size of the cell (in CC method). A small value of N (a small cell size for CC method) results in a spiky density field which makes it vulnerable to unrealistic density values due to Poisson noise and random clustering of spatially uncorrelated galaxies. A large value of N (a large cell size for CC method) is prone to underestimation of the surface density and over-smoothing the details of galaxy distribution. Also, for the nearest neighbor method, the sum of the area (volume) assigned to each galaxy is not equal to the total area (volume) of any survey and it has also been shown that its integral over all area (volume) diverges.\\
More importantly, it is still not clear whether the environmental effects depend on the physical scales at which the environment is estimated. For example, \cite{Kauffmann04}, \cite{Blanton06} and \cite{Cucciati10} find that the effect of environment on the star formation history of galaxies is only effective at the small scales ($\lesssim$1 Mpc) whereas \cite{Balogh04} find that the equivalent width of H$\alpha$ (a measure of the specific star formation rate in galaxies) is a function of environment measured on scales of $\sim$1 and $\sim$5 Mpc, independent of each other. Scale-dependent density estimators such as the NN and CC algorithms are not able to directly address this issue.\\
In this work, we perform a comprehensive analysis of the (surface) density field using the weighted versions of the 5$^{th}$ NN, 10$^{th}$ NN, adaptive kernel smoothing, Voronoi tessellation and Delaunay triangulation methods. This is done using a K$_{s}$-band selected sample of galaxies in the COSMOS field \citep{Scoville07}. The large size of the COSMOS ($\sim$2 deg$^{2}$), together with the high accuracy of the photometric redshifts in this field, enable us to delineate the Large Scale Structure with great accuracy out to z$\sim$3. In this analysis, we use the full photo-z probability distribution function (PDF) of individual galaxies to significantly reduce the projection effect. The performance of each method is checked with extensive realistic and Monte-Carlo simulations. We then apply the algorithm to a mass complete sample of galaxies to investigate the role of density-based definition of environment on the rest-frame color evolution of quiescent galaxies, as well as the evolution of comoving number and mass density of massive systems out to z$\sim$3.\\
The format of this paper is as follows. In Section \ref{data}, we briefly review the data used to estimate the density field. Section \ref{method} outlines the algorithm used to determine the density field, followed by the surface density estimation methods in Section \ref{denfield}. Section \ref{sim} deals with the simulations used to check the performance of different density estimators. Comparisons between different methods are given in Section \ref{compare}. In Section \ref{science}, we study the color evolution of quiescent galaxies, as well as number and mass density of massive systems as a function of environment, using a stellar mass complete sample from the COSMOS field. We give a summary of this work in Section \ref{concl}.\\
Throughout this work, we assume a flat concordance $\Lambda$CDM cosmology with H$_{0}$=70 kms$^{-1}$ Mpc$^{-1}$, $\Omega_{m}$=0.3 and $\Omega_{\Lambda}$=0.7. All magnitudes are expressed in the AB system and stellar masses are given assuming a Chabrier IMF.

\section{Data \& Sample selection} \label{data}

In order to examine different density estimation techniques, we select an area of $\sim$1.8 deg$^{2}$ in the COSMOS field \citep{Scoville07,Capak07}. Here, we use the COSMOS UltraVISTA K$_{s}$-band selected photometric redshift (photo-z) catalog \citep{McCracken12,Ilbert13}. The K$_{s}$-band selection is equivalent to a stellar mass-selected sample and enables us to reliably detect galaxies (especially quiescent systems) at intermediate to high redshifts. This catalog consists of ground and space based photometric data in 30 bands, spanning the range UV to mid-IR. Using this extensive data set, accurate photometric redshifts are measured for galaxies out to z$\sim$3. Stars, AGNs and X-ray sources were identified and removed from the sample. The final selection criteria used to define the sample are:
\begin{itemize}
\item An area of 149.3$<\alpha_{2000}<$150.8 and 1.6$<\delta_{2000}<$2.8 corresponding to $\sim$1.8 deg$^{2}$. The area is selected to be large enough for the effect of LSS to be discernible.
\item Redshift within the range 0.05$<$z$<$3.2 to allow photometric redshifts with high degree of accuracy.
\item Galaxies brighter than K$_{s}<$24, since the photo-z uncertainties increase significantly at K$_{s}\gtrsim$24. 
\end{itemize}
The total sample contains 191151 galaxies.
 
\section{Determination of the Large Scale Distribution of Galaxies} \label{method}

In any study of the LSS of galaxies, we need the angular position ($\alpha$,$\delta$) of galaxies and a measure of their radial distance (using spectroscopic or photometric redshifts) to estimate the density field. Here, we use the full photo-z probability distribution function (PDF) of each individual galaxy to significantly reduce the projection effect. In summary, the density field estimation involves the following steps:
\begin{enumerate}
\item First, we construct a series of overlapping redshift slices (z-slices). The widths of these slices are obtained by the photo-z accuracy of the underlying sample of galaxies at each redshift.
\item For each z-slice, we select a \it subset \rm of galaxies belonging to that slice. The galaxies in the subset are selected in such a way that the median of their photo-z PDF is located within that z-slice. For a given subset, we associate a weight to each galaxy. This weight shows the likelihood of each galaxy belonging to the z-slice of our interest. The weight is determined by the photo-z PDF of each galaxy and the boundaries of the z-slice.
\item At each z-slice and for its subset of galaxies, we then compute the weighted surface density field using one of the estimation methods described in Section \ref{denfield}.
\end{enumerate}
In the following, we give a detailed description of points 1 \& 2. Using this information, we then apply different density estimators to the data in Section \ref{denfield}.
\begin{figure}
  \centering
  \includegraphics[width=3.5in]{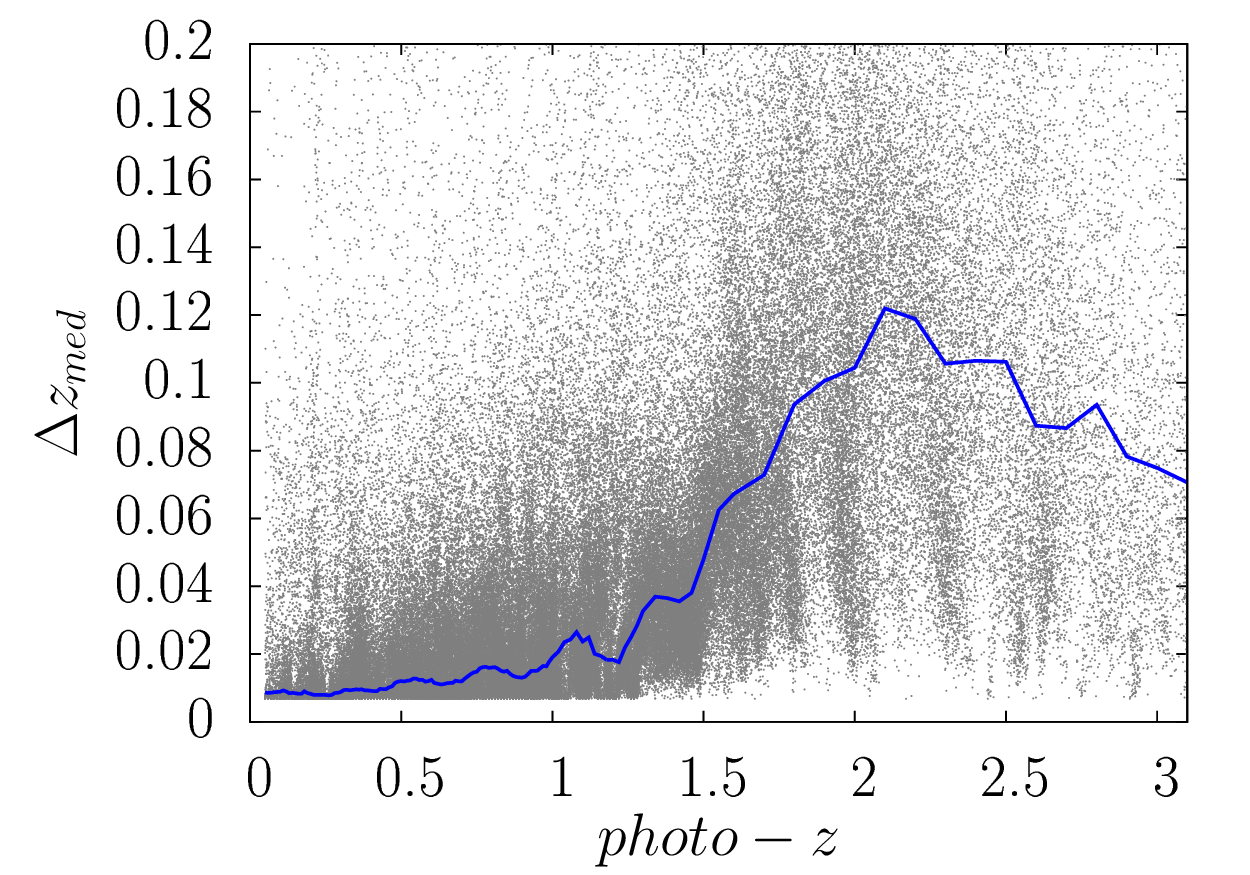}
\caption{ The photo-z uncertainty of galaxies used in this study. The blue solid line shows the median of the photo-z uncertainty ($\Delta$z$_{med}$) at each redshift. $\Delta$z for each galaxy is calculated as the (lower \& higher) 68\% confidence interval of its photo-z PDF. Note the accuracy of photo-z values. The median of the photo-z uncertainty is $\Delta$z$_{med}\lesssim$0.01 out to z$\sim$1, reaching $\Delta$z$_{med}\sim$0.1 at z$\sim$2.}
\label{fig:delta-z}
\end{figure}

\subsection{Redshift Slicing} \label{z-slice}

In order to construct the density field of galaxies, one needs a large, homogeneous and unbiased sample of galaxies with very accurate redshifts. However, there are serious complications in constructing such samples. Firstly, building a homogeneous sample is difficult. Secondly, relying on a purely spectroscopic sample would bias the study as it would only concentrate on bright objects and mostly those with emission lines (i.e., star-forming galaxies which are often less clustered compared to quiescent systems). Spectroscopic samples also do not usually target enough number of galaxies in dense regions. Thirdly, finding accurate photometric redshifts requires a homogeneous set of multi-waveband photometric data.  
Thanks to the wealth of the photometric data available in the COSMOS field, we have been able to obtain highly accurate photometric redshifts for a large population of galaxies in this field \citep{Mobasher07,Ilbert09,Ilbert13}. Figure \ref{fig:delta-z} shows the median of photo-z uncertainty ($\Delta$z$_{med}$) as a function of (photometric) redshift for the sample of galaxies used in this study. $\Delta$z for each galaxy is calculated as the (lower \& higher) 68\% confidence interval of its photo-z PDF. The median of the photo-z uncertainty is $\Delta$z$_{med}\lesssim$0.01 out to z$\sim$1, reaching $\Delta$z$_{med}\sim$0.1 at z$\sim$2. Studying the 3D density estimations solely based on the photo-z of galaxies, without taking their photo-z uncertainties into account, may result in an erroneous evaluation of the density field. This is mainly due to the fact that the physical lengths that correspond to even the smallest photo-z uncertainties are larger than the typical sizes of most physical structures such as galaxy clusters and groups (e.g. at z$\sim$1, the photo-z uncertainty of $\Delta$z$\sim$0.01 is equivalent to the comoving radial length of $\sim$24 Mpc). Here, we limit our analysis to the 2D surface density estimations by constructing a series of narrow slices in the redshift space. This is done by defining a series of overlapping redshift slices (z-slices). The width of each slice is set by the photometric redshift accuracy of the data at any given redshift. The slices overlap to allow proper contribution from galaxies which reside close to the boundaries of each slice. In practice, we make sure that approximately half of each z-slice is trespassed into its adjacent z-slices. The width of each z-slice ($\delta$z) at redshift z is defined as twice the median of the photo-z uncertainty ($\Delta$z$_{med}$) at that redshift: 
\begin{equation}
\delta z=2\Delta z_{med}
\end{equation} 
Figure \ref{fig:delta-z} shows how $\Delta$z$_{med}$ changes with redshift. In this study, a total number of 133 overlapping z-slices are used over the redshift range 0.1$<$z$<$3.2.\\      
Choosing a much narrower z-slice results in smaller number statistics and decreased completeness of the underlying sample of galaxies. This subsequently leads to an underestimation of the surface density field. On the other side, broadening the z-slice increases the risk of contamination from foreground and background galaxies as well as a possible overestimation of the density field. Thus, our choice is a compromise between both limits.

\subsection{Photo-z Probability Weights} \label{weight}
\begin{figure*}
  \centering
  \includegraphics[width=7.0in]{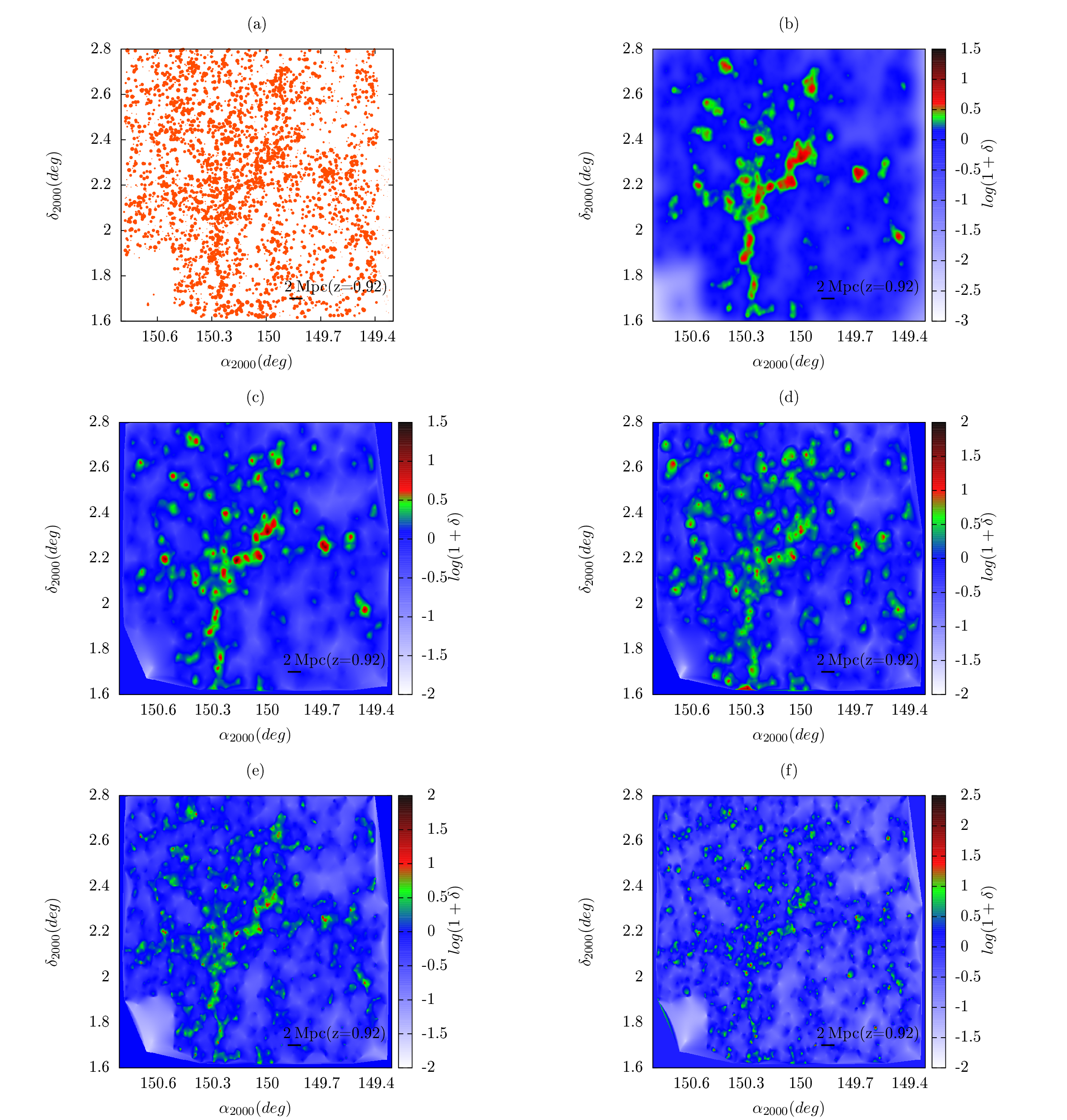}
\caption{ Comparison of surface density fields in the COSMOS at z=0.92, estimated with different methods. (a) Distribution of galaxies for a z-slice located at z=0.92. The size of each point is proportional to the galaxy weight. (b) Surface density field estimated using the weighted adaptive kernel smoothing method for the distribution of galaxies at z=0.92. (c) Surface density field estimated using the weighted 10$^{th}$ NN method (d) 5$^{th}$ NN (e) Voronoi tessellation and (f) Delaunay triangulation methods.}
\label{fig:all-092}
\end{figure*}

To estimate the surface density field, we use the full information of the photo-z PDF for individual galaxies. First, for each z-slice, we select a subset of galaxies. Galaxies in each subset are selected in such a way that their median photo-z PDF is located within the boundaries of that z-slice. We assign a weight (w$_{i}$) to each galaxy in this subset based upon its photo-z PDF. This probability weight is calculated by measuring what percentage of the photo-z PDF of each galaxy is contained within the boundaries of the considered z-slice. This weight represents the likelihood that the galaxy in question resides within that z-slice. In practice, it is done by intersecting the z-slice of our interest with the photo-z PDF of each galaxy in the subset. In order to enhance the computational performance, we assume a Gaussian photo-z PDF for galaxies except for those that have a second photo-z peak with P$>$5\% in their PDF. For the latter, the full photo-z PDF is used in estimating the weights. Eventually, at each redshift and for each z-slice, we have a subset of galaxies with associated weights that are later used as the input for the weighted surface density estimation methods. Figure \ref{fig:all-092} (a) shows the distribution of galaxies for a z-slice centered at z=0.92 in the COSMOS. For demonstration, the size of each point is selected to be proportional to the galaxy weight. This subset of galaxies with their associated weights is used to estimate the weighted surface density field at z=0.92.\\
\begin{table}
\begin{center}
{\scriptsize
{{Table 1: Overdensity values evaluated using the adaptive kernel estimator for sample galaxies selected in Section \ref{data}.}} 
\begin{tabular}{lcccccc}
\hline
\noalign{\smallskip}
ID & $\alpha_{2000}$ & $\delta_{2000}$ & photo-z & K$_{s}$ & log($\Sigma$/$\Sigma_{median}$)$_{kernel}$\\
 & deg & deg & & mag & \\
\hline
1 & 150.02025 & 1.67951 & 0.0532 & 22.69 & -0.5811\\
2 & 150.39144 & 1.68383 & 0.0506 & 23.30 & -0.6667\\
3 & 149.58068 & 1.68829 & 0.0512 & 23.71 & -0.4306\\
\hline
\end{tabular}
\label{table:overdens}
}
\caption{Notes: Table 1 is published in its entirety in the electronic edition of the {\it Astrophysical Journal}. A portion is shown here for guidance regarding its form and content.}
\end{center}
\end{table}
We note that the weight introduced in this section can be generalized to other galaxy properties such as stellar mass, luminosity and color, depending on the astrophysics of our problem.   
        
\section{Surface Density Estimation: The Methods} \label{denfield}

Using the information obtained in Section \ref{method}, we now construct the density field for galaxies in the COSMOS, using weighted versions of four independent methods: adaptive kernel, nearest neighbor, Voronoi tessellation and Delaunay triangulation (Sections \ref{kernel} to \ref{delaunay}). We then evaluate their performance with simulations in Section \ref{sim} and compare them with each other in Section \ref{compare}.

\subsection{Weighted Adaptive Kernel Estimator} \label{kernel}

In this method, we use an iterative procedure to compute the surface density field \citep{Darvish14}. First, we estimate the surface density associated with the $i$th galaxy in a given z-slice, $\hat{\Sigma}_i$, by summing over all the weighted fixed kernels placed on the positions of galaxies, $j$, where $i\neq j$:
\begin{equation} 
\hat{\Sigma}_i= \frac{1}{\sum_{\substack{j=1\\j\neq i}}^{N} w_j}\sum_{\substack{j=1\\j\neq i}}^{N} w_j K(\bf r\rm_i,\bf r\rm_j,h)
\end{equation}
where N is the number of galaxies in the subset, $K(\bf r\rm_i,\bf r\rm_j,h)$ is the fixed kernel, $\bf r\rm_i$ is the position of the galaxy for which the initial estimate of surface density is measured and $\bf r\rm_j$ is the position of the rest of the galaxies. The width of the kernel is expressed by the parameter, h, which is a proxy for the degree of smoothing. For the first estimate of the density, this is taken to be fixed. For the kernel smoothing function, we use a 2D symmetric Gaussian defined as:
\begin{equation}
K(\bf r\rm_i,\bf r\rm_j,h)=\frac{1}{2\pi h^2}e^{-\frac{|\bf r\rm_i-\bf r\rm_j|^2}{2h^2}}
\end{equation}
A large kernel width (h) results in over-smoothing of the density field which tends to wash out real features while a small value tends to break up regions into smaller uncorrelated substructures. Here, we use a fixed physical length of h=0.5 Mpc which corresponds to the typical value of $R_{200}$ for X-ray clusters \& groups in the COSMOS field \citep{Finoguenov07,George11}. However, a constant value of h for the whole field has the problem that it underestimates the surface density in crowded regions while overestimates in sparsely populated areas. To overcome this problem, we introduce 
adaptive smoothing width, h$_i$,  which is a measure of the local surface density associated with each galaxy, $\hat{\Sigma}_i$. This is defined as h$_i$=h$\times\lambda_i$, where $\lambda_i$ is a parameter that is inversely proportional to the square root of the surface density associated with the $i$th galaxy, at the position of that galaxy \citep{Silverman86}:
\begin{equation}
\lambda_{i}=(G/\hat{\Sigma}(\bf r\rm_i))^{0.5}
\end{equation}
Where G is the geometric mean of all the $\hat{\Sigma}(\bf r\rm_i)$ values.
Having the adaptive kernel, we now calculate the surface density field, $\Sigma({\bf r})$, on each location on a fine 2D grid, ${\bf r}$=(x,y) as:
\begin{equation} 
\Sigma(\bf r\rm)= \frac{1}{\sum_{i=1}^{N} w_i}\sum_{i=1}^{N} w_i K(\bf r\rm,\bf r\rm_i,h_i)
\end{equation}
The surface density field is evaluated on a fine grid with a grid size (resolution) of 50 Kpc at each redshift. We note that the surface density field estimated through this method is almost independent of the type of the kernel function. We examined several other standard kernels including exponentially decaying, Epanechnikov (K(r)$\propto$(1-r$^{2}$)) and cosine arch (K(r)$\propto$cos($\frac{\pi}{2}$r)) and did not find any significant difference in the final results. Throughout this work, we define overdensity as:
\begin{equation}
1+\delta= \frac{\Sigma}{\Sigma_{median}}
\end{equation}
where $\Sigma_{median}$ is the median of the surface density field at each redshift. Figure \ref{fig:all-092} (b) shows an example of the surface density field for a z-slice in the COSMOS field centered at z=0.92, using the weighted adaptive kernel density estimator. Note the wide range of overdensities and the variety of environments, including dense regions linked together through thread-like filamentary structures.\\
In table 1, we have provided an example of the overdensity values estimated with the adaptive kernel smoothing method for our sample galaxies (Section \ref{data}). K$_{s}$ magnitudes and photo-z estimates are extracted from publicly available catalogs of \citep{McCracken12} and \citep{Ilbert13}, respectively. We recommend using galaxies that are far from the edge of the field and/or the masked regions. A catalog of overdensity values with other density estimation methods is also available upon request. 
 
\subsection{Weighted k-Nearest Neighbors(k-NN) estimator} \label{NN}

In a regular (non-weighted) k-NN method, the inverse of the area containing the kth 
nearest neighbors to each galaxy is used as a proxy for local surface density at the position of that galaxy. Since galaxies in our sample are weighted, we incorporate the role of weights in the density estimation. The surface density at the position of the $i$th galaxy, $\Sigma({\bf r\rm_{i}})$, is estimated as:    
\begin{equation}
\Sigma(\bf r\rm_{i})= \frac{\sum_{j=1}^{k} jw_{ij}}{\pi \sum_{j=1}^{k} w_{ij}d_{ij}^{2}}
\end{equation}
Where $\bf r\rm_{i}$ is the position of $i$th galaxy, w$_{ij}$ is the photo-z probability weight (Section \ref{weight}) associated with the $j$th nearest neighbor to the galaxy located at $\bf r\rm_{i}$ and d$_{ij}$ is the distance between them (the distance between the galaxy positioned at $\bf r\rm_{i}$ and its $j$th neighbor). Using the distance to the fifth (k=5) or tenth (k=10) nearest neighbor is very common in literature (see for example, \cite{Cooper05,Sobral11}). In this work, we use the weighted versions of both the fifth (NN$_{5}$) and tenth (NN$_{10}$) nearest neighbor estimators. Figures \ref{fig:all-092} (c) and (d) show examples of the surface density field at z=0.92, estimated using NN$_{10}$ and NN$_{5}$, respectively. The NN$_{5}$ method gives a larger density value in very dense regions compared to NN$_{10}$, as expected. The overall traces of NN$_{5}$ are seen in NN$_{5}$ plot. However, the NN$_{5}$-based density field looks spikier and clumpier due to the smaller physical lengths that the NN$_{5}$ method spans.  
      
\subsection{Weighted Voronoi Tessellation Estimator} \label{voronoi}

In a simple Voronoi tessellation method, we divide the 2D z-slice plane into a number of regions (Voronoi cells) assigned to each galaxy located at $\bf r\rm_{i}$. The Voronoi cell of each galaxy is defined as all points in the z-slice plane that are closer to that galaxy than to any other galaxy. It is acquired from the intersection of half-planes. Based on this definition, in more crowded, denser regions, the Voronoi cells of galaxies incline towards smaller values, therefore, we can use the inverse of the area of the Voronoi cell of each galaxy as a measure of the local density at the position of that galaxy \citep{Ebeling93,Bernardeau96}. This is given by:
\begin{equation}
\Sigma(\bf r\rm_{i})= \frac{1}{A_{i}}
\end{equation}
Where $\Sigma({\bf r\rm_{i}})$ is the surface density at the position of the $i$th galaxy and A$_{i}$ is the area of its Voronoi cell.\\
In order to assimilate the role of galaxy weights, we tried two different approaches.
In the first approach, we modified the metric used for the definition of distance between points in the z-slice plane and galaxies. This was done by dividing the Euclidean metric by the weight of each galaxy. Implementing this approach to the data was not successful and resulted in unrealistically high density values for the majority of galaxies (with small weights) located in sparsely populated regions. Thus, we alternatively use a Monte-Carlo acceptance-rejection process in order to take the role of weights into consideration. The steps are as follows:
\begin{enumerate}
\item For each galaxy with its corresponding weight w$_{i}$, we generate a random number R$_{i}$ between the minimum and maximum weight values in each z-slice.
\item If w$_{i}$ $>$ R$_{i}$, we accept the galaxy (with its associated weight w$_{i}$) in the density estimation procedure.
\item We estimate the surface density for the accepted galaxies using a simple Voronoi tessellation method. This surface density is evaluated at the position of accepted galaxies only.
\item We interpolate\footnote[1]{We use natural neighbor interpolation developed by Sibson.} the estimated densities into the points of a grid ($\bf r$) (the grid resolution is 50 Kpc at each redshift) constructed on the z-slice plane. This gives us a Monte-Carlo estimated surface density field $\tilde{\Sigma}({\bf r})$.  
\item We repeat the above procedure N times and take the mean of all the Monte-Carlo density fields as the actual density field $\Sigma(\bf r)$: 
\begin{equation}
\Sigma(\bf r\rm)=\frac{1}{N}\sum_{m=1}^{N} \tilde{\Sigma}_{m}({\bf r})
\end{equation}
\end{enumerate}
Finally, the local surface density of each galaxy is then estimated as that of its closest point in the grid constructed over the z-slice plane. To save computational time, we use N=10 in step 5. Figure \ref{fig:all-092} (e) shows an example of the surface density field at z=0.92, estimated with the weighted Voronoi tessellation algorithm. Note the large dynamical range of overdensities. Unlike the nearest neighbor, Voronoi tessellation is scale-independent and is able to span a wide range of physical lengths. Also, it does not make any assumptions about the geometry and morphology of the structures in the density field. This characteristic makes it superior to adaptive kernel and nearest neighbor methods.   

\subsection{Weighted Delaunay Triangulation Estimator} \label{delaunay}

This method relies on segmenting the z-slice plane into triangles whose vertices are 
defined by the position of galaxies in the z-slice plane. For each triangle, these three vertices (position of galaxies) are selected such that their circumcircle does not encompass any other galaxy. In this method, each galaxy is eventually surrounded by a series of neighboring triangles whose overall area tend to be smaller in denser regions \citep{Schaap00,Platen11}. For the $i$th galaxy surrounded by $m$ neighboring triangles, the estimated surface density is expressed as the inverse of the sum of the areas of its neighboring triangles:
\begin{equation}
\Sigma(\bf r\rm_{i})= \frac{1}{\sum_{n=1}^{m} a_{n}}
\end{equation}
Where $\Sigma({\bf r\rm_{i}})$ is the surface density at the position of the $i$th galaxy and a$_{n}$ is the area of the $n$th triangle neighboring the $i$th galaxy.\\
In order to take the weight of galaxies into account, we utilize a Monte-Carlo acceptance-rejection approach explained in Section \ref{voronoi}. Figure \ref{fig:all-092} (f) shows an example of the surface density field at z=0.92, estimated with the weighted Delaunay method. When compared with other estimation methods, this method overestimates the densities in very dense regions and the resulting density field is clumpier and contains much more substructures.
 
\section{Simulations} \label{sim}

In order to evaluate the performance of each surface density estimation method, we run two sets of simulations. The details are given in the following subsections.

\subsection{Simulation 1} \label{sim1}

In the first set of simulations, we apply the surface density estimators to a sample of galaxies randomly drawn from some previously known surface density profiles \citep{Scoville07b}. Since the simulated surface density values are known, we can directly compare them with the estimated surface densities that are predicted by each of our density estimators. Here, we make 30 different simulated structures. These structures are placed on an area similar to that of the COSMOS field. 
\begin{figure*}
  \centering
  \includegraphics[width=6.2in]{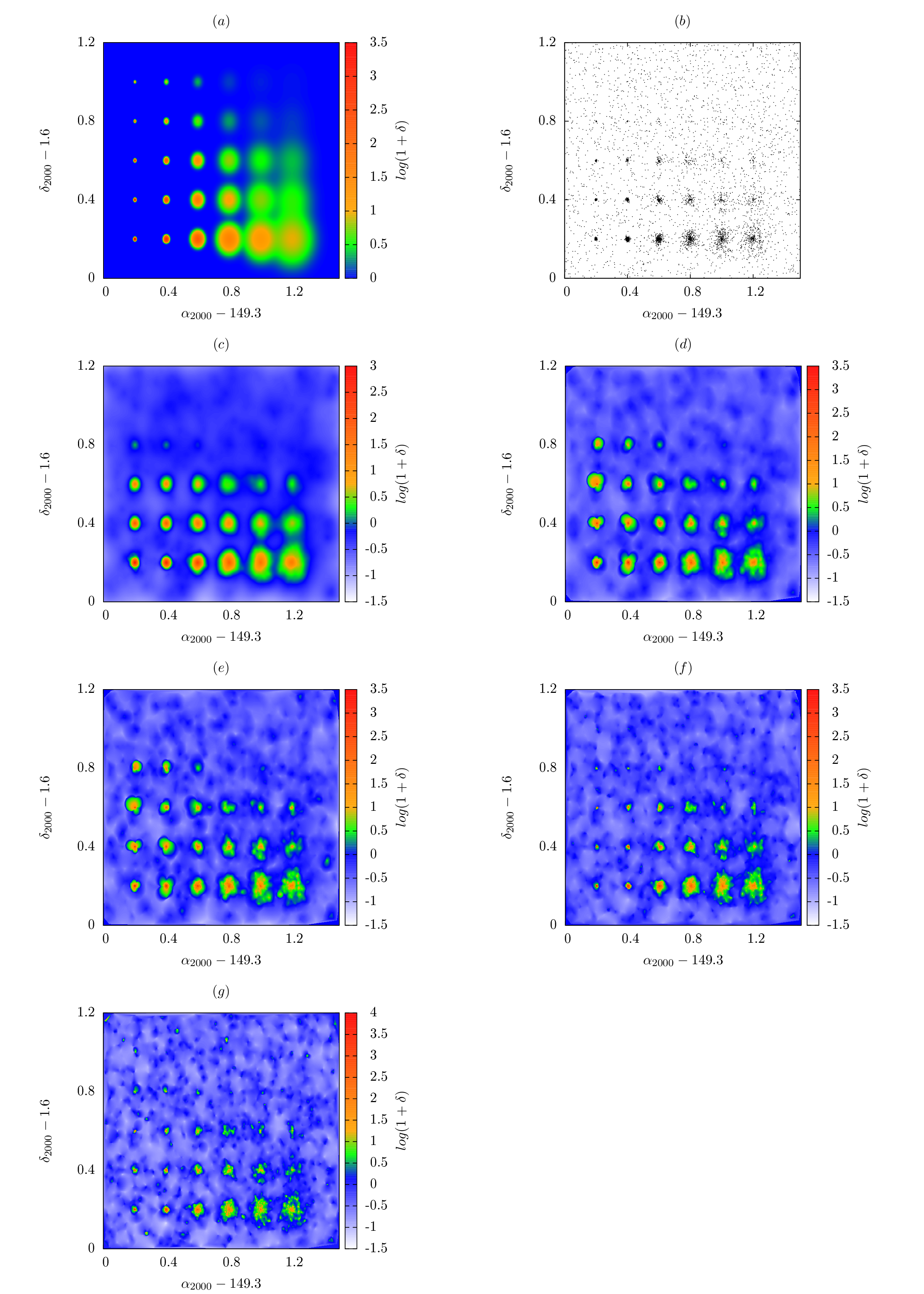}
\caption{(a) Overdensity map for simulated structures. Here, we make 30 different simulated structures placed on a constant background. They are assumed to be located at z=1 and cover an area similar to that of the COSMOS field. The structures have Gaussian profiles with a variety of sizes (0.1-2 Mpc) and the number of galaxies in these structures is in the range 3-300. (b) Distribution of galaxies randomly drawn from the structures. The simulated structures and the field contain 5000 galaxies. Almost half of them (2778) belong to the structures while the rest are randomly distributed on an area covering 1.5$\times$1.2 deg$^{2}$, which serve like the field. (c) Predicted overdensity maps using the adaptive kernel, (d) 10$^{th}$ NN, (e) 5$^{th}$ NN, (f) Voronoi tessellation and (g) Delaunay triangulation. All these methods perform relatively well when compared to the expected density field, with adaptive kernel and Voronoi tessellation performing relatively better than other estimators.}
\label{fig:sim1}
\end{figure*}
The simulated structures and the field contain 5000 galaxies at z =1, similar to the total number of observed galaxies at that redshift in the COSMOS field. For simplicity, we assume all to have the same weight (w=1). 2222 of these galaxies are randomly distributed on an area covering 1.5$\times$1.2 deg$^{2}$. The rest of them are drawn randomly from a series of Gaussian profiles. These Gaussian structures have a variety of sizes (0.1-2 Mpc) and the number of galaxies in these structures is in the range 3-300. The properties of these structures are shown in table 2. We apply the density field estimation methods explained in Section \ref{denfield} to the simulated galaxies to explore how well they can predict the input density field. Figure \ref{fig:sim1} shows the density field of the simulated structures, the distribution of galaxies randomly drawn from the structures and the predicted density fields using different estimation methods. All these methods perform relatively well when compared to the expected density field. We stress that some of the difference between the expected and predicted density fields is due to the shot noise, which becomes specifically important when the number of galaxies in simulated structures is small and the width of the profile is large. In order to quantify the performance of different density estimation methods, we define the mean squared error (MSE) as:
\begin{equation}
MSE=\frac{1}{N}\sum_{i=1}^{N} (\Delta_{i}^{sim}-\Delta_{i})^2
\end{equation}
Where N is the total number of generated galaxies, $\Delta_{i}^{sim}$=log(1+$\delta_{i}^{sim}$) is the logarithm of the overdensity of the $i$th galaxy in the simulation and $\Delta_{i}$ is a similar quantity, predicted by one of the density estimator methods. A smaller value of MSE shows a higher similarity between the simulation and predicted density fields and is a measure of the performance of the estimation methods in predicting the true value of the density field. The MSE values are tabulated in table 3. It is clear that the adaptive kernel, Voronoi tessellation and NN$_{10}$ methods outperform the NN$_{5}$ and Delaunay triangulation algorithms.\\
We note that due to the catastrophic failure in estimating the photometric redshifts, the actual discrepancy between the simulated and predicted density fields can be higher.However, the catastrophic failure fraction (defined as the fraction of galaxies which satisfy $|$z$_{phot}$-z$_{spec}|$/(1+z$_{spec}$)$>$0.15) for K$_{s}<$24 galaxies in the COSMOS field is relatively small and estimated to be $<$1\% for bright sources, reaching $\lesssim$10\% for fainter galaxies \citep{Ilbert13}. We redo the simulation taking into account the catastrophic failure and estimate the new MSE statistics. We assume that 10\% of simulated galaxies are affected by the catastrophic failure (i.e., they do not actually belong to the simulated z-slice) and they are randomly distributed over the area containing the simulated structures (random distribution is a fair assumption since catastrophic failure mostly affects fainter galaxies which tend to be less clustered). This means that the actual background for the overdensity map of simulated structures should have a smaller value. With the new simulation, all density field estimators perform almost equally worse with catastrophic failure (as predicted). The MSE statistics is increased by $\sim$0.01-0.02 ($\sim$3-6\%) for all density estimators. In this work, we do not correct for the catastrophic failure but we highlight that it has a small effect and the trends are not affected much by it.\\      
\begin{table}
\begin{center}
{\scriptsize
{{Table 2: Properties of simulated structures. The structures have a Gaussian profile with a variety of sizes and number of galaxies.}} 
\begin{tabular}{lccccc}
\hline
\noalign{\smallskip}
Structure & $\alpha_{2000}$ & $\delta_{2000}$ & $\sigma$ & N\\
& deg & deg & Mpc & \\
\hline
1 & 149.5 & 1.8 & 0.1 & 300\\
2 & 149.7 & 1.8 & 0.2 & 300\\
3 & 149.9 & 1.8 & 0.5 & 300\\
4 & 150.1 & 1.8 & 1.0 & 300\\
5 & 150.3 & 1.8 & 1.5 & 300\\
6 & 150.5 & 1.8 & 2.0 & 300\\
7 & 149.5 & 2.0 & 0.1 & 100\\
8 & 149.7 & 2.0 & 0.2 & 100\\
9 & 149.9 & 2.0 & 0.5 & 100\\
10 & 150.1 & 2.0 & 1.0 & 100\\
11 & 150.3 & 2.0 & 1.5 & 100\\
12 & 150.5 & 2.0 & 2.0 & 100\\
13 & 149.5 & 2.2 & 0.1 & 50\\
14 & 149.7 & 2.2 & 0.2 & 50\\
15 & 149.9 & 2.2 & 0.5 & 50\\
16 & 150.1 & 2.2 & 1.0 & 50\\
17 & 150.3 & 2.2 & 1.5 & 50\\
18 & 150.5 & 2.2 & 2.0 & 50\\
19 & 149.5 & 2.4 & 0.1 & 10\\
20 & 149.7 & 2.4 & 0.2 & 10\\
21 & 149.9 & 2.4 & 0.5 & 10\\
22 & 150.1 & 2.4 & 1.0 & 10\\
23 & 150.3 & 2.4 & 1.5 & 10\\
24 & 150.5 & 2.4 & 2.0 & 10\\
25 & 149.5 & 2.6 & 0.1 & 3\\
26 & 149.7 & 2.6 & 0.2 & 3\\
27 & 149.9 & 2.6 & 0.5 & 3\\
28 & 150.1 & 2.6 & 1.0 & 3\\
29 & 150.3 & 2.6 & 1.5 & 3\\
30 & 150.5 & 2.6 & 2.0 & 3\\
\hline
\end{tabular}
\label{table:sim}
}
\end{center}
\end{table}

\subsection{Simulation 2} \label{sim2}

In the previous simulations, although the ``true" density values are known (which make the comparison with the predicted densities feasible), the complex geometry of the real astronomical density fields and their diverse dynamical range are not fully considered; i.e, the simulated density profiles have simple mathematical shapes whereas the real astronomical data consist of an irregular web of filaments, voids, walls and clusters with a variety of physical scales and geometries. Furthermore, the previous simulation greatly suffers from the shot noise, especially when the number of randomly drawn galaxies is small and/or the known density profiles have broad, close-to-the-field shapes. In order to consider the complexity of the cosmic web in the analysis, one can use the numerical simulations which produce mock galaxy distributions that resemble the real density maps. Since the ``true" density field is unknown in these quasi-real simulations, we should rely on an estimator to have an initial guess of the density field. In the second set of simulations, we make a sample of galaxies that resemble the true distribution of galaxies in the COSMOS. This is performed using a Monte-Carlo acceptance-rejection approach, a method similar to the one explained in Section \ref{delaunay}, taking the following steps:
\begin{enumerate}
\item For each z-slice, we estimate the surface density field for the real data using one of the density estimators explained in Section \ref{denfield}. For simplicity, we assume that all galaxies in the z-slice have the same weight (w=1). The estimated density field gives us a first order approximation of the shape of the true density field.
\item For each z-slice, we randomly select a position, (x,y)$_{random}$, in the area covering the data and assign a random number, $\Sigma_{random}$, between the minimum and the maximum density values in that z-slice to the point (x,y)$_{random}$. We also report the density value at this random position from step 1 and call it $\hat{\Sigma}$.
\item If $\hat{\Sigma}$$>$$\Sigma_{random}$, we accept the point (x,y)$_{random}$ as one of the points in the simulated dataset and consider $\hat{\Sigma}$ as its true density. 
\item For each z-slice, we repeat steps 1-3 until we make a simulated dataset with the same number of galaxies in the actual data set.
\end{enumerate}
\begin{table}
\begin{center}
{\scriptsize
{{Table 3: Comparison between the simulated density field (simulation 1) and the one predicted by different estimators. We use the MSE measure for comparison. A smaller value of MSE indicates a better performance.}} 
\begin{tabular}{lcccccc}
\hline
\noalign{\smallskip}
 & Kernel & NN$_{10}$ & NN$_{5}$ & Voronoi & Delaunay\\
\hline
MSE & 0.26 & 0.28 & 0.30 & 0.27 & 0.34\\
\hline
\end{tabular}
\label{table:MSE1}
}
\end{center}
\end{table}
\begin{table}
\begin{center}
{\scriptsize
{{Table 4: Comparison between the simulated density field (simulation 2) and the one predicted by different estimators. We use the MSE measure for comparison. A smaller value of MSE indicates a better performance.}} 
\begin{tabular}{lcccccc}
\hline
\noalign{\smallskip}
 & Kernel & NN$_{10}$ & NN$_{5}$ & Voronoi & Delaunay\\
\hline
MSE & 0.003 & 0.035 & 0.062 & 0.025 & 0.158\\
\hline
\end{tabular}
\label{table:MSE2}
}
\end{center}
\end{table}
Given the above steps, for each z-slice, we now have a sample of simulated galaxies with \it known \rm surface densities. The distribution of these galaxies resembles the actual data in the COSMOS. We apply our surface density estimation methods to the simulated data with known densities and compare their expected and predicted densities. Although now, we consider the complexity of real galaxy distribution into account, these simulations rely on an initial estimator to determine the rough shape of the density field (i.e., step 1). We therefore use the adaptive kernel density estimator in order to perform the first step of the simulation \footnote[2]{This results in a bias in favor of the performance of the adaptive kernel density estimator when we compare it with simulations. However, we used other estimators as well to initiate the simulation and found that the adaptive kernel estimator has an overall relatively good performance in all these simulations.}. Table 4 shows the performance of different surface density estimators when compared with the new set of simulations. Here, we use the MSE measure introduced in Section \ref{sim1} to compare the results. According to the new simulations and the results of the MSE statistics, the adaptive kernel, Voronoi tessellation and NN$_{10}$ methods perform better in estimating the simulated density field. In both sets of simulations, the NN$_{5}$ and (especially) the Delaunay tessellation algorithms seem to fall behind compared to other estimators. 
\begin{figure}
  \centering
  \includegraphics[width=3.5in]{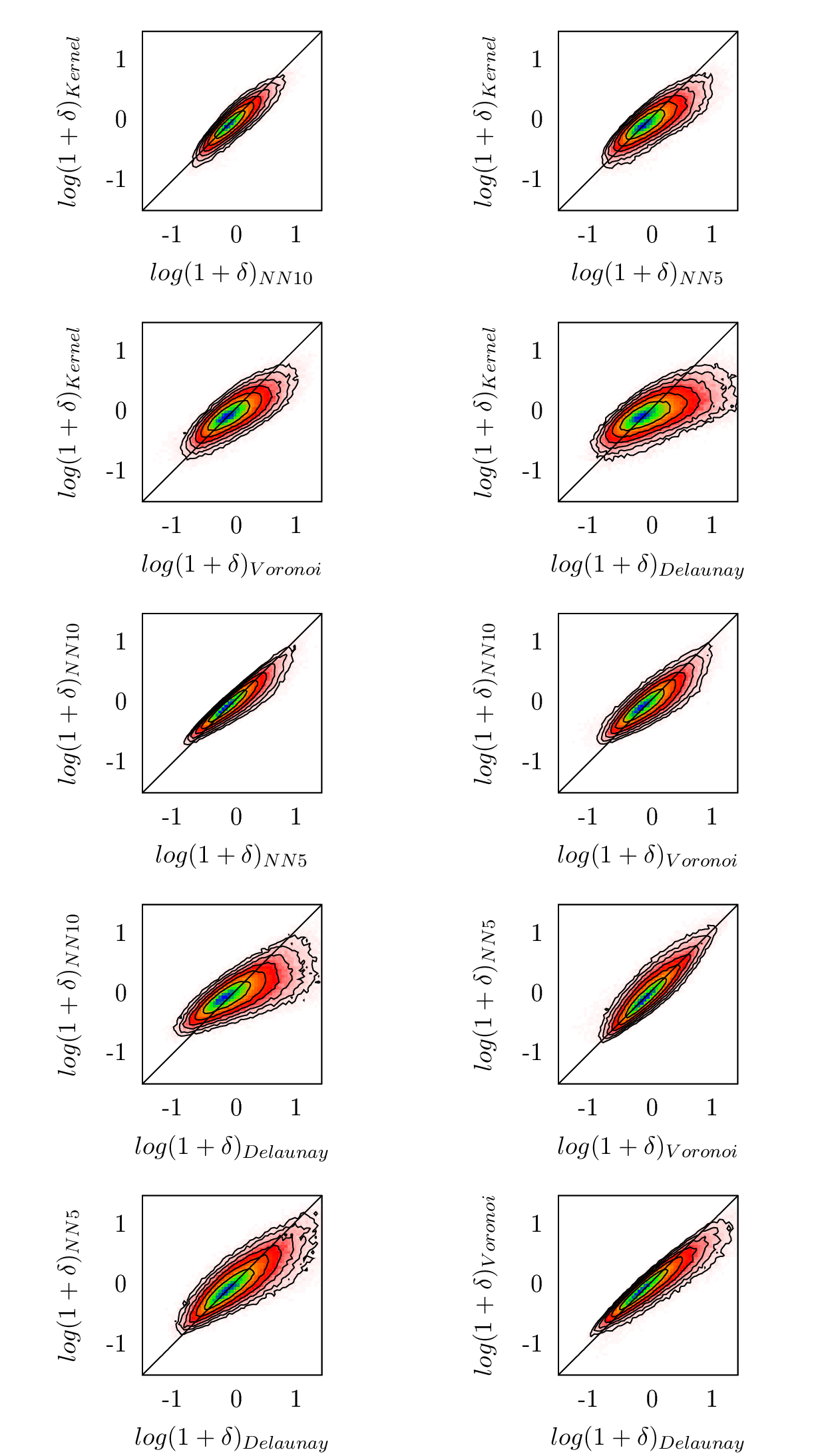}
\caption{Comparison between different estimation methods using 176893 galaxies at 0.1$<$z$<$3.2. In order to minimize the edge effect, we limit the comparison to galaxies that are 1 Mpc away from the edge of the field and large masked regions. There is a good agreement between different estimation methods over $\sim$2 orders of magnitude in overdensity values. However, when we compare the Delaunay triangulation method with the rest, it overestimates the density values in dense regions while underestimates the density in sparsely populated areas.}
\label{fig:comparison}
\end{figure}

\begin{figure*}
  \centering
  \includegraphics[width=7.0in]{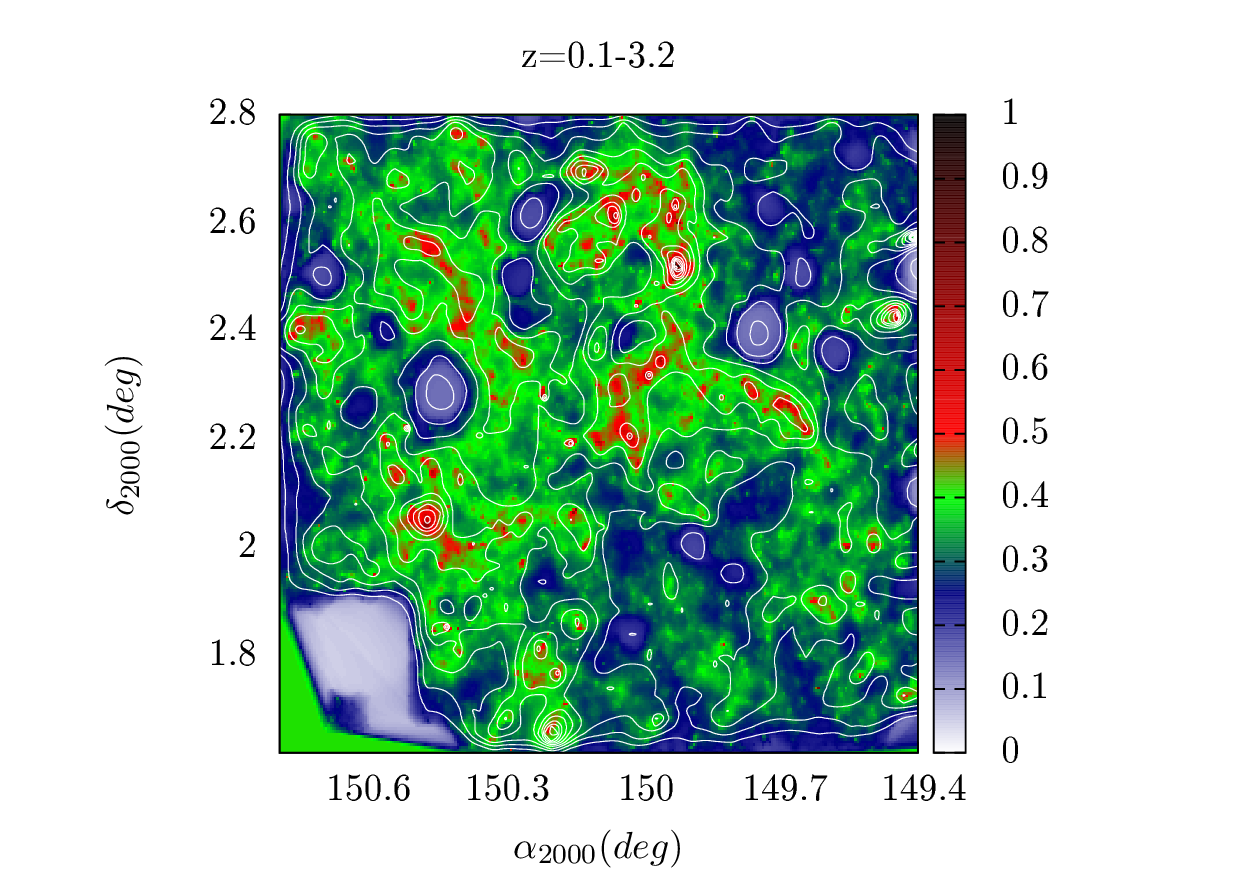}
\caption{Projected overdensity map in the whole COSMOS field at 0.1$<$z$<$3.2, using the weighted Voronoi tessellation method. In constructing this, the overdensity maps from all the z-slices between z=0.1-3.2 are stacked and normalized to the peak. We compare it with the stacked overdensity map from Voronoi-based algorithm of \cite{Scoville13}. Contours are used to demonstrate the map from \cite{Scoville13}. Contours are at levels 0.2 to 1 with 0.05 spacing between levels. There is a very good agreement between \cite{Scoville13} and our work. We also find a relatively good agreement between the denser regions in our work and the position of X-ray clusters/groups \citep{Finoguenov07,George11}, as well as the projected mass map form weak lensing analysis of \cite{Massey07} in the COSMOS.}
\label{fig:density-all}
\end{figure*}

\section{Comparison} \label{compare}

Using two sets of simulations, we showed that all density estimation methods perform relatively well in estimating the density field. However, the adaptive kernel, Voronoi tessellation and NN$_{10}$ are found to outperform the others. In this section, we compare different estimation methods together regardless of their performance. Due to the edge effect, the surface density of galaxies located near the edge of the survey and masked areas are unrealistically  underestimated. In order to take this into account, we limit the comparison to galaxies that are more than 1 Mpc (physical) away from the edge of survey and large masked areas. Figure \ref{fig:comparison} shows a comparison between different estimation methods using 176893 galaxies at 0.1$<$z$<$3.2. There is a relatively good agreement between different estimation methods over $\sim$2 orders of magnitude in overdensity values. However, when we compare the Delaunay triangulation method with the rest, it overestimates the density values in dense regions while underestimates the density in sparsely populated regions (field) when compared with adaptive kernel and NN$_{10}$ methods.\\
Each surface density estimator method has its own advantages and disadvantages as explored below:
\begin{itemize}
\item Adaptive Kernel: Since we smooth the density field with a suitable kernel function, this method is less affected by the shot noise and possible random clustering of foreground and background sources. Calculation of the global kernel width (h) is motivated based on some physical scales (e.g. the typical size of galaxy clusters and groups) which makes it suitable for practical observational situations. We can easily add weights to the estimator by multiplying the kernel function with a proper weight. This method conserves the total number of galaxies; i.e, the integral of the surface density field over the whole area yields the total number of galaxies. However, the selection of the appropriate kernel size is a serious problem, as a small kernel size tends to overestimate the sparsely populated regions in the density field whereas a large kernel size washes out real features. This is partly overcome by adaptively changing the kernel width, at the expense of increasing the computational time by adding an extra step to initially estimate the density field at the position of galaxies. Since we estimate the surface density at the positions of a grid, an intermediate interpolation is required to assign the density values from the grid points to the position of galaxies. It is also computationally expensive in its adaptive format as it requires extra steps to determine the local adaptive kernel width.
\item k-NN: It is by far the easiest method to implement and fastest to perform computationally. It estimates the surface density directly at the position of galaxies which does not require any interpolation. Adding weights to this method is readily done. However, its performance strongly depends on the number of neighbors considered in the analysis (k). A small value of k results in a spiky density field which makes it vulnerable to unrealistic density values due to Poisson noise and random clustering of spatially uncorrelated galaxies. A large value of k is prone to underestimation of the surface density and oversmoothing the details of galaxy distribution. The sum of the area assigned to each galaxy is not equal to the total area of survey. It has also been shown that its integral over all area diverges.
\item Voronoi Tessellation: This method covers a wide dynamical range in densities and is able to estimate densities in a broad range, from the dense core of clusters to sparse regions devoid of galaxies. It is a non-parametric and scale-independent method which makes no prior assumption about the shape of the density field. We directly evaluate the surface density at the position of galaxies. However, we cannot assign closed Voronoi areas to galaxies near the edge of the field. Adding weights to this method is not straightforward. Here, we used a Monte-Carlo method to take the role of weights into consideration. However, this comes at the expense of a computationally expensive process by making several Monte-Carlo samples. Apart from its computational time, it is a robust estimator.
\item Delaunay Tessellation: In terms of advantages and disadvantages, it is very similar to the Voronoi tessellation method. However, we can assign delaunay area to all galaxies in the sample even those that are located at the edge of the field. Also in this method, the total area assigned to galaxies surpasses the area of the survey. Despite its similarity to the Voronoi tessellation, this method does not perform well compared to other estimation methods (at least for the COSMOS field). 
\end{itemize}     
Due to the good performance of the Voronoi tessellation method, its large dynamical range, scale independence and no prior assumption it makes about the morphology of the structures, we use it for the scientific analysis in the next sections. Figure \ref{fig:density-all} shows the projected overdensity map in the whole COSMOS field at 0.1$<$z$<$3.2, using the weighted Voronoi tessellation method. In constructing this, we stack the overdensity maps from all the z-slices between z=0.1-3.2 and normalize it to the peak of stacked overdensities. We compare it with the stacked overdensity map from Voronoi-based algorithm of \cite{Scoville13}. Contours are used to demonstrate the map from \cite{Scoville13}. Contours are at levels 0.2 to 1 with 0.05 spacing between levels. In terms of sample selection, our work is similar to that of \cite{Scoville13}. We see a very good agreement between \cite{Scoville13} and our work. We also find a relatively good agreement between the denser regions in our work and the position of X-ray clusters/groups \citep{Finoguenov07,George11}. When compared with the projected mass map form weak lensing analysis of \cite{Massey07} in the COSMOS field, our projected map agrees with that of \cite{Massey07}. We are able to recover all the massive structures in the weak lensing map. Our estimated density field is also consistent quantitatively with the density field of \cite{Kovac10} who used zCOSMOS spectroscopic data out to z$\sim$1, the optical galaxy groups of \citep{Knobel09,Knobel12} using zCOSMOS data set and the protocluster candidates of \cite{Chiang14} at 1.6$<$z$<$3.1 in the COSMOS field. In the following sections, we use the weighted Voronoi tessellation method to study the dependence of the observable parameters on environment.

\begin{table*}
\begin{center}
{\scriptsize
{{Table 5: Properties of mass complete samples used in this study}} 
\begin{tabular}{lcccc}
\hline
\noalign{\smallskip}
Sample Number & Redshift Range & Mass Completeness limit & Number of galaxies\\
              &                & log($M_{\odot}$) &\\
\hline
1 & 0.1$\leq$z$<$0.5 & 9.14 & 9338\\
2 & 0.5$\leq$z$<$0.8 & 9.47 & 11760\\ 
3 & 0.8$\leq$z$<$1.1 & 9.70 & 13885\\
4 & 1.1$\leq$z$<$1.5 & 9.93 & 13640\\
5 & 1.5$\leq$z$<$2.0 & 9.97 & 12217\\ 
6 & 2.0$\leq$z$<$3.1 & 9.97 & 12641\\ 
\hline
\end{tabular}
\label{table:Mass-complete}
}
\end{center}
\end{table*}

\section{Dependence of the Physical Properties of Galaxies on the Environment} \label{science}

Using the surface density field constructed in the COSMOS field, we now study the dependence of the observable parameters of galaxies on their local density. We first need to define a mass complete sample of galaxies at different redshifts. In the following sections, we select the sample first and then use it to investigate the role of environment on properties of galaxies. 

\subsection{Sample Selection}

In the following subsections, we first define a mass complete sample at different redshifts. We also explain the selection of quiescent/star-forming galaxies.
 
\subsubsection{Stellar Mass Complete Samples} \label{mass-comp}

The magnitude cut K$_{s}<$24 introduced in Section \ref{data} results in a magnitude-limited sample whose mass completeness is a function of redshift and stellar M/L ratio. In order to estimate the mass completeness of our sample, we use the method explained in \cite{Pozzetti10} and \cite{Ilbert13}. First, for each galaxy, we calculate the limiting stellar mass (M$_{lim}$); i.e., the mass it would have at its redshift, if its apparent magnitude were equal to the magnitude limit of the sample (K$_{s}$=24). This is given by log(M$_{lim}$/M$_{\odot}$)=log(M/M$_{\odot}$)+0.4(K$_{s}$-24), where M is the estimated stellar mass of the galaxy with apparent magnitude K$_{s}$. At each redshift, the stellar mass completeness limit (M$_{comp}$) corresponds to the mass with 95\% of the galaxies having their M$_{lim}$ below the stellar mass completeness limit. 
\begin{figure}
  \centering
  \includegraphics[width=3.5in]{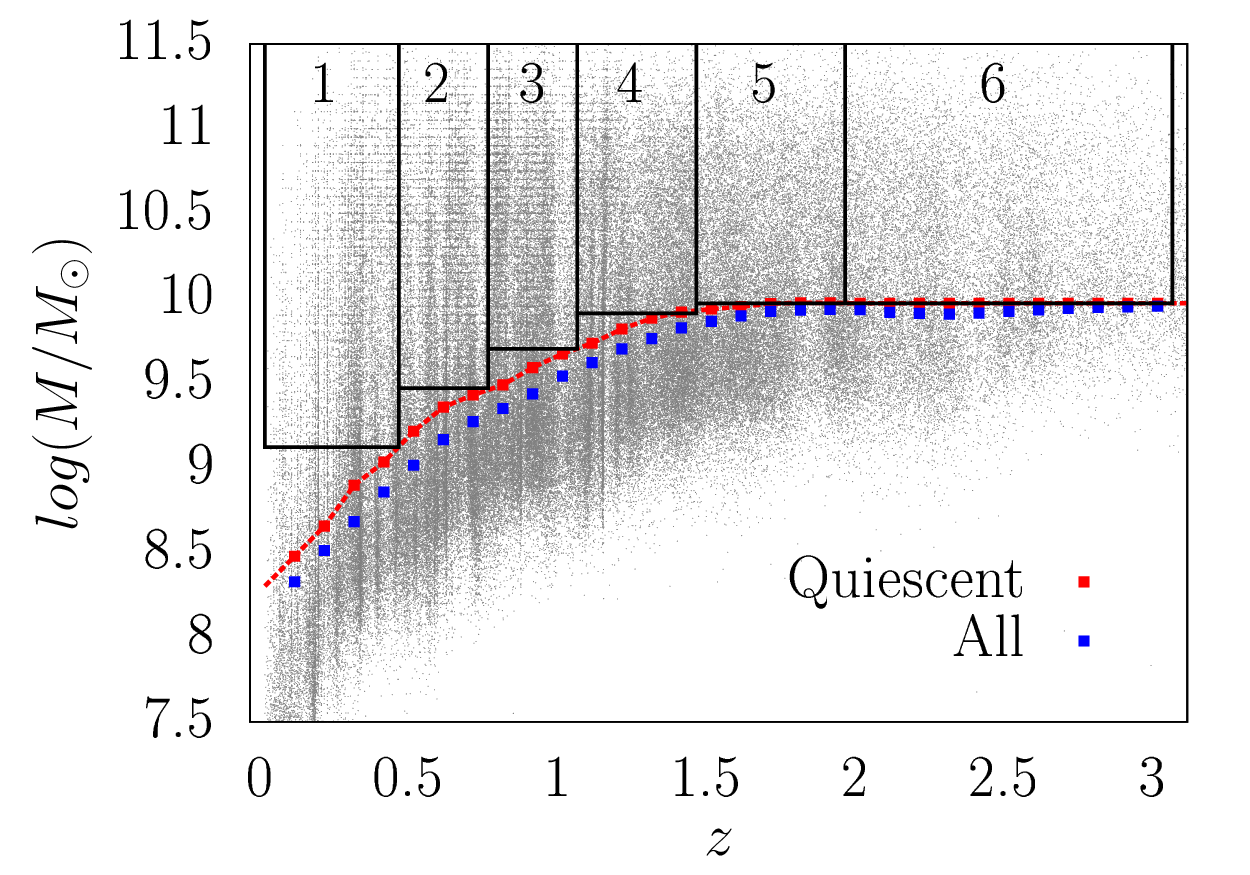}
\caption{Mass completeness limit for all the galaxies, along with that of the quiescent systems as a function of redshift. Mass completeness limit is defined in such a way that only less than 5\% of galaxies could be missed in the lower mass distribution of galaxies. Using the mass completeness limits for the quiescent systems, we define six mass complete samples at different redshifts, as shown with labels 1-6 here.}
\label{fig:mass-comp}
\end{figure}
This guarantees that only less than 5\% of galaxies could be missed in the lower mass regime of the stellar mass distribution of galaxies. The mass completeness limit also depends on the stellar M/L ratio and is higher for quiescent galaxies. Therefore, in constructing the mass complete samples at each redshift, we make sure that we rely on the quiescent galaxies to estimate the stellar completeness limit. This minimizes the loss of low-mass quiescent galaxies especially at higher redshifts. The selection of quiescent/star-forming galaxies is explained in Section \ref{NUV-R-J}. Figure \ref{fig:mass-comp} shows the stellar mass of galaxies as a function of redshift for magnitude-limited sample (K$_{s}<$24) defined in Section \ref{data}. The mass completeness limits for all the galaxies (quiescent \& star-forming) and quiescent systems are shown in Figure \ref{fig:mass-comp}. As mentioned already, only less than 5\% of galaxies could be missed in the lower mass population of galaxies. The mass completeness limit is higher for quiescent systems. Here, we define six mass complete samples out to z$\sim$3, as shown in Figure \ref{fig:mass-comp}. The properties of these mass complete samples are given in table 5.
\begin{figure}
  \centering
  \includegraphics[height=3.5in,width=3.5in]{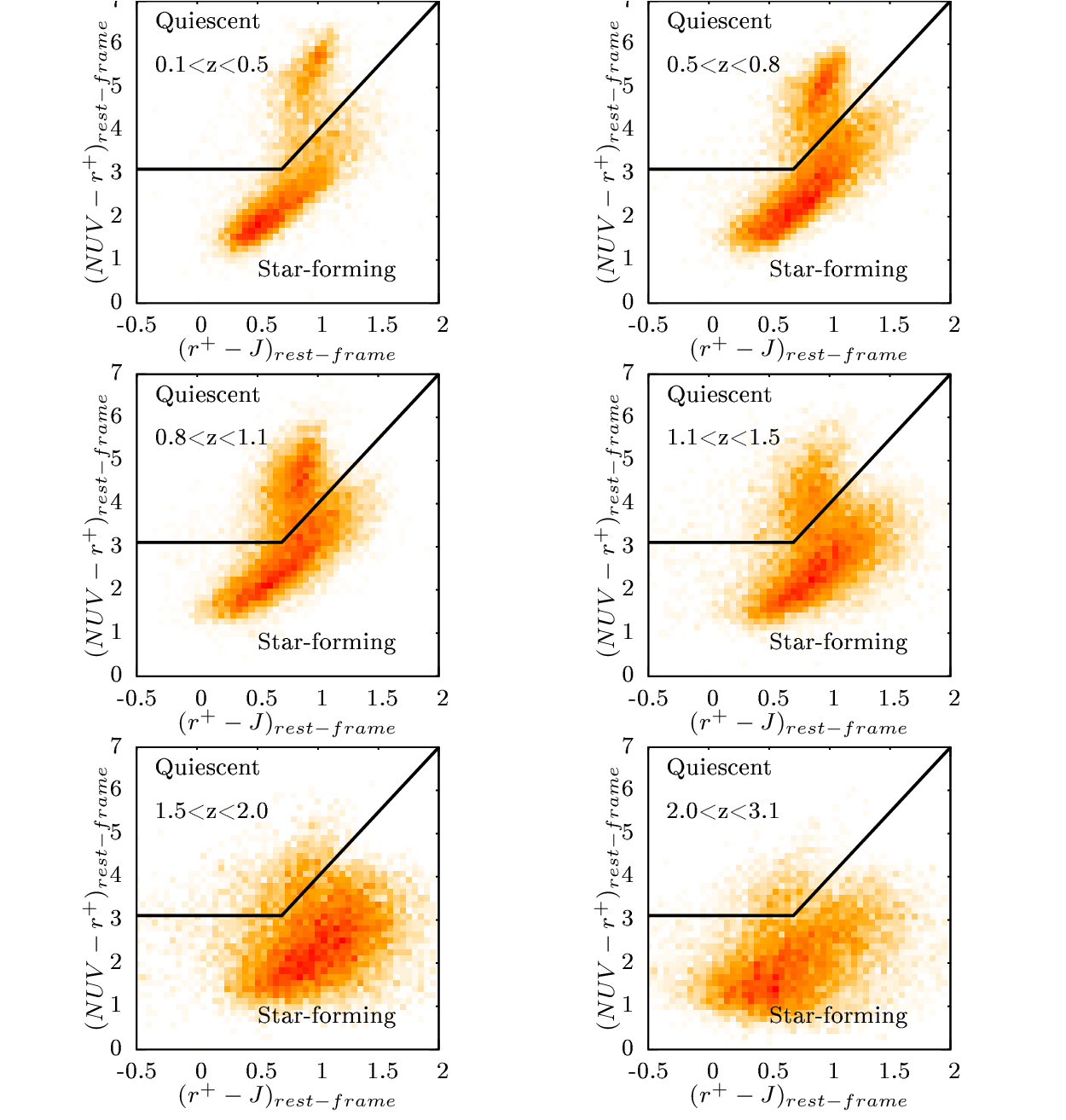}
\caption{ NUV-$r^{+}$ versus $r^{+}$-J color-color plots used to select quiescent and star-forming populations in our mass complete samples at z=0.1-3.1. Galaxies with their rest-frame color NUV-$r^{+}$ $>$ 3.1 and NUV-$r^{+}$ $>$ 3($r^{+}$-J)+1 are selected as quiescent systems \citep{Ilbert13}.}
\label{fig:nuv-r-j}
\end{figure}
\subsubsection{Selection of Quiescent/Star-forming Systems} \label{NUV-R-J} 
 
The selection of quiescent and star-forming systems can be performed using a single rest-frame color through the Color-Magnitude Diagram (CMD). However, the single color selection is problematic for several reasons. The existence of dusty, star-forming galaxies that mimic the color of quiescent systems can significantly contaminate the quiescent population and tend to unrealistically increase the quiescent fraction and their comoving number densities \citep{Brammer09}. The larger scatter in the rest-frame color at higher redshifts (mostly due to photo-z uncertainties) can wash out the red sequence and the disappearance of the rest-frame color bimodality at z$\gtrsim$1.5 \citep{Williams09}. Here, we use rest-frame two-color NUV-$r^{+}$ versus $r^{+}$-J in order to select quiescent and star-forming populations in our mass complete samples. It has been shown that the rest-frame NUV-$r^{+}$ color is a better indicator of the recent star formation activity \citep{Martin07} and has a wider dynamical range compared to the more commonly used rest-frame U-V color \citep{Ilbert13}. Here, galaxies with their rest-frame color NUV-$r^{+}$ $>$ 3.1 and NUV-$r^{+}$ $>$ 3($r^{+}$-J)+1 are selected as quiescent systems \citep{Ilbert13}. Figure \ref{fig:nuv-r-j} shows the rest-frame NUV-$r^{+}$ versus $r^{+}$-J distribution of galaxies in the mass complete samples and the color cuts used to separate quiescent and star-forming systems. Two distinct populations of galaxies are clearly seen out to z$\sim$3. We stress that in our two-color selection technique, adding dust to the star-forming galaxies causes them to move diagonally from the bottom left to the top right of Figure \ref{fig:nuv-r-j}, making them separable from the quiescent systems \citep{Williams09,Ilbert13}. 

\subsection{Evolution of Rest-frame Color as a Function of Environment For Quiescent Galaxies} \label{CMEZ}
\begin{figure*}
  \centering
  \includegraphics[width=7.0in]{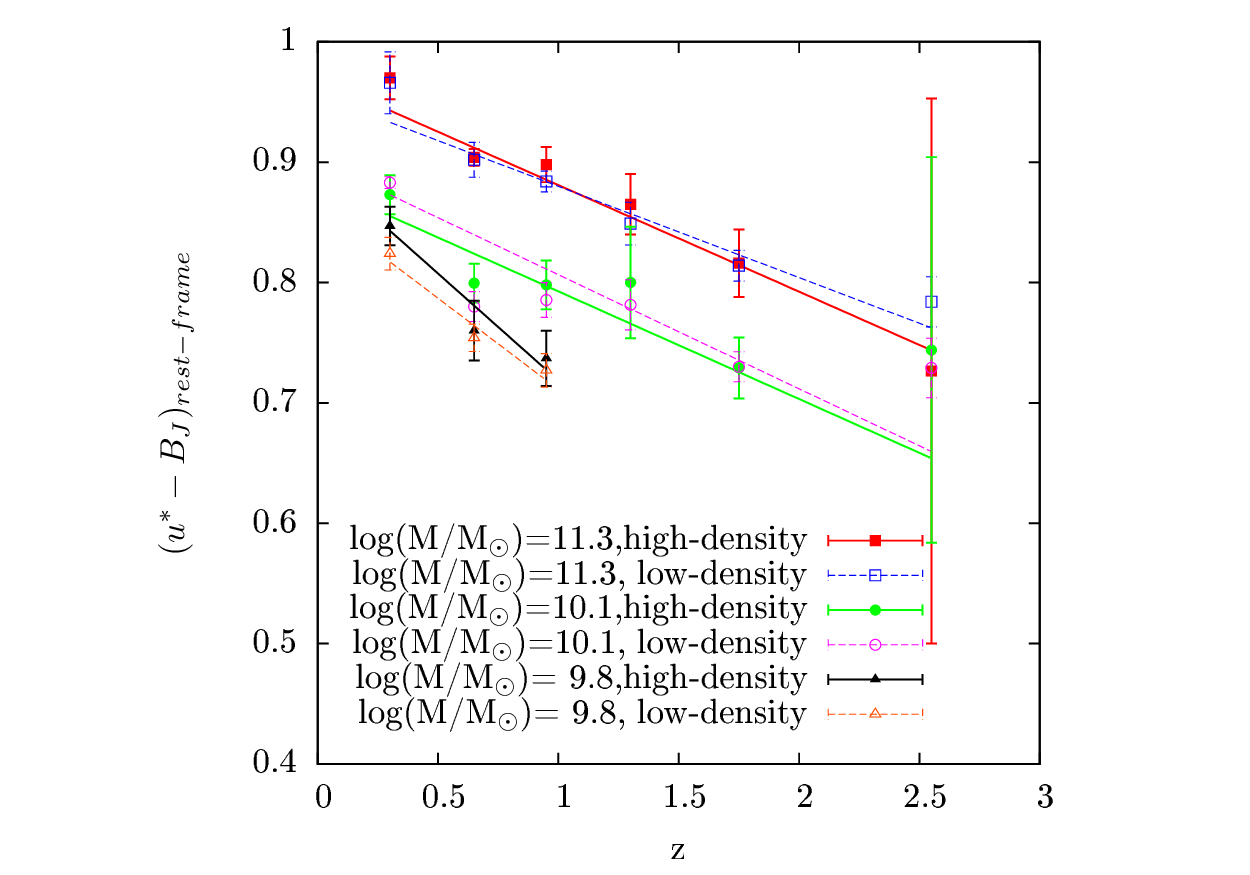}
\caption{Redshift evolution of the median rest-frame u$^{*}$-B$_{J}$ color for quiescent galaxies with different stellar masses located in high and low-density environments. For clarity, the evolution is shown for only a few stellar masses. The color uncertainties are estimated using 10000 bootstrap resamples added in quadrature to the median observational uncertainties in color. We assume that the evolution of rest-frame u$^{*}$-B$_{J}$ color for quiescent systems is linear with redshift and fit a straight line to the median colors at any given mass and environment. At a fixed redshift and stellar mass, the color of quiescent galaxies is independent of environment. However, at a fixed redshift and environment, the color of quiescent galaxies depends on stellar mass. Quiescent galaxies become redder with cosmic time and their evolution is independent of the environment they reside in. Quiescent galaxies more massive than log(M/M$_{\odot}$)$\gtrsim$10 become $\sim$0.2 mag redder in rest-frame u$^{*}$-B$_{J}$ color since z$\sim$2.5. Since z$\sim$1, less massive systems (log(M/M$_{\odot}$)$\sim$9.5-10) redden by $\sim$0.1 mag, regardless of their environment. We also find that more massive quiescent galaxies (log(M/M$_{\odot}$)$\sim$11) are $\sim$0.1 mag redder in the rest-frame u$^{*}$-B$_{J}$ color compared to less massive (log(M/M$_{\odot}$)$\sim$10) systems and this color difference is almost independent of the environment and redshift.}
\label{fig:color}
\end{figure*}
We now investigate the effect of environment on the rest-frame u$^{*}$-B$_{J}$ color of quiescent systems with different stellar masses and its evolution with redshift. We note that in selecting the quiescent population, we do not use the single rest-frame u$^{*}$-B$_{J}$ color due to the issues expressed in Section \ref{NUV-R-J}. The selection is alternatively done based on rest-frame NUV-$r^{+}$ versus $r^{+}$-J plot. Here, we define two environments: high-density environment (galaxy group and cluster scales) is defined as regions with overdensity values log(1+$\delta$)$\geq$0.5 and low-density environment (field-like environment) is defined as those with log(1+$\delta$)$<$0.5. The selection of the cut between low and high-density environments is somewhat arbitrary. However, it is shown (Darvish et al. 2015 in prep.) that the environmental effects (e.g., the increase in the fraction of quiescent galaxies with overdensity) start to effectively rise at log(1+$\delta$)$\gtrsim$0.5 in the COSMOS field. Throughout this work, we use this cut to separate the density field into low- and high-density environments. Based on this definition, we study the redshift evolution of the rest-frame u$^{*}$-B$_{J}$ color as a function of environment and stellar mass. Figure \ref{fig:color} shows the redshift evolution of the median rest-frame u$^{*}$-B$_{J}$ color for quiescent galaxies with different stellar masses located in high- and low-density environments. For clarity, the evolution is shown for only a few stellar mass bins ($\Delta$M=$\pm$0.1 dex around the selected masses). We choose the center of the redshift bins (table 5) as the redshift of the given points. The median color at each given environment, stellar mass and redshift is estimated using all the quiescent systems that are in that environment, with their redshift located in the redshift range of the mass complete samples and with stellar masses within $\Delta$M$\pm$0.1 dex of the given stellar mass. The color uncertainties are estimated using 10000 bootstrap resamples added in quadrature to the median observational uncertainties in color. We assume that the evolution of the rest-frame u$^{*}$-B$_{J}$ color for quiescent systems is linear with redshift and fit a straight line to the median colors at any given mass and environment:
\begin{equation}
(u^{*}-B_{J})_{rest}(z)=\alpha z+(u^{*}-B_{J})_{rest}(z=0)
\end{equation}
The result of the fit to the median colors for different environments and stellar masses is given in table 6. According to Figure \ref{fig:color} and table 6, quiescent galaxies become redder with cosmic time and their evolution is independent of the environment they reside in. We particularly find that irrespective of environment, quiescent galaxies more massive than log(M/M$_{\odot}$)$\gtrsim$10 become $\sim$0.2 mag redder since z$\sim$2.5. This is in agreement with \cite{Kriek08} who showed a reddening of $\sim$0.25 mag in the rest-frame U-B color for massive (2$\times$10$^{11}$ M$_{\odot}$) red galaxies in the field since z$\sim$2.3. This reddening is also seen in less massive quiescent systems (log(M/M$_{\odot}$)=10-11) and is independent of environment since z$\sim$3. Due to the incompleteness in stellar mass, we can not study the color evolution of less massive galaxies (log(M/M$_{\odot}$)$\lesssim$10) at z$\gtrsim$1. However, at z$\lesssim$1, these systems (log(M/M$_{\odot}$)$\sim$9.5-10) redden by $\sim$0.1 mag, regardless of their environment. 
We also see that at any given redshift out to z$\sim$3, more massive quiescent systems are redder compared to less massive systems, regardless of their environment. Particularly, we find that more massive quiescent galaxies (log(M/M$_{\odot}$)$\sim$11) are $\sim$0.1 mag redder in the rest-frame u$^{*}$-B$_{J}$ color compared to less massive (log(M/M$_{\odot}$)$\sim$10) systems and this color difference is independent of the environment. The fact that more massive quiescent systems are redder than the less-massive ones is a manifestation of the Color-Magnitude Relation (CMR) since the stellar mass is proportional to the luminosity (magnitude) of galaxies. The CMR is seen in both the field (less dense) and cluster (dense) quiescent galaxies, and in the local universe out to higher redshifts (see for example, \citep{Bower92,Blakeslee03,Bell04,Mei09,Wilson09,Brammer09,Papovich10}).\\
For quiescent galaxies, we find that the rest-frame color at a fixed environment depends on stellar mass but at a fixed mass is independent of environment. At first glance, this seems to contradict the studies showing a tight correlation between color and local density of galaxies (for example, \citealp{Balogh04}). However, the stellar mass also depends strongly on density and color-density relation is actually a manifestation of a more fundamental color-mass relation. When controlled for the stellar mass (or luminosity), the color-density relation becomes independent of the environment (overdensity).\\ 
\begin{table}
\begin{center}
\scriptsize
{
{{Table 6: Parameters of the linear fit ((u$^{*}$-B$_{J}$)$_{rest}$(z)=$\alpha$z+(u$^{*}$-B$_{J}$)$_{rest}$(z=0)) of the rest-frame u$^{*}$-B$_{J}$ color evolution for quiescent galaxies with different stellar masses.}}
\begin{tabular}{lccccc}
\noalign{\smallskip}
\hline
\noalign{\smallskip}
M & \multicolumn{2}{c}{$\alpha$} & \multicolumn{2}{c}{(u$^{*}$-B$_{J}$)$_{rest}$(z=0)} \\
\cline{2-3} \cline{4-5}
log(M$_\odot$) & high-density & low-density & high-density & low-density \\
\hline
11.3 & -0.089$\pm$0.020 & -0.076$\pm$0.010 & 0.969$\pm$0.016 & 0.956$\pm$0.012 \\
10.8 & -0.082$\pm$0.032 & -0.076$\pm$0.011 & 0.917$\pm$0.026 & 0.920$\pm$0.016 \\
10.5 & -0.082$\pm$0.023 & -0.065$\pm$0.016 & 0.916$\pm$0.024 & 0.882$\pm$0.019 \\
10.1 & -0.089$\pm$0.020 & -0.095$\pm$0.022 & 0.882$\pm$0.018 & 0.901$\pm$0.018 \\ 
9.8 & -0.178$\pm$0.041 & -0.151$\pm$0.035 & 0.896$\pm$0.025 & 0.863$\pm$0.024 \\ 
9.6 & -0.176 & -0.157 & 0.885 & 0.853 \\
\hline
\end{tabular}
\label{table:line}
}
\end{center}
\end{table}
The environmental independence of color-mass relation is consistent with several previous studies. In the local universe, For example, \cite{Balogh04} showed that at a fixed luminosity (equivalent to a fixed stellar mass), the mean rest-frame color of red galaxies in the SDSS at z$<$0.08 is nearly independent of their environment. Independently, \cite{Baldry06} showed that at z$<$0.1, the color-mass distribution of galaxies (red \& blue) does not significantly change with environment. The same result is also seen in \cite{Hogg04} who found the independence of red bulge-dominated galaxy colors on environment in the SDSS at z$<$0.12. Using the SDSS data, \cite{Park07} showed that when morphology and luminosity (equivalent to stellar mass in our study) are fixed, other physical properties of galaxies, including color is nearly independent of local density of galaxies. They also showed that more luminous galaxies are redder than less luminous systems and this is independent of the environment. This is consistent with our results. The environmental invariance of color for bright galaxies in the SDSS is also seen in \cite{Tanaka04}.\\
Several studies at higher redshifts are also consistent with our results. For example, \cite{Cassata07} showed that at z$\sim$0.7 in the COSMOS field, the observed color-magnitude diagram for red and early-type galaxies is independent of the local density of galaxies. \cite{Scodeggio09} found that for a sample of galaxies in the VVDS survey at 0.2$<$z$<$1.4, color is independent of the local density of galaxies at a fixed stellar mass. Using the zCOSMOS data out to z$\sim$1, \cite{Moresco10} found that the rest-frame color of red galaxies at a fixed stellar mass is almost independent of environment but at a fixed environment, depends on the stellar mass. They also found that the average colors of massive red galaxies (log(M/M$_{\odot}$)=10.8) are redder than low-mass galaxies ((log(M/M$_{\odot}$)=10) throughout their entire redshift range. This is entirely consistent with our results in this section. By extending this study to higher redshifts, we demonstrate that the results hold to z$\sim$3. \cite{Cucciati10} showed that for red massive galaxies ((log(M/M$_{\odot}$)$\gtrsim$10.7) in the zCOSMOS 10k-sample, the color-density relation is globally flat up to z$\sim$1, consistent with our results. Recently, \cite{Bassett13} compared the rest-frame U-B color of galaxies in a cluster with those in the field at z=1.6 in the CANDELS-UDS field and found no difference between the color of quiescent galaxies in these two environments. This agrees well with our results at z$>$1.

\subsection{Effect of the Environment on the Comoving Number and Mass density of Massive Galaxies}

In this section, we investigate the evolution of comoving number (n) and mass ($\rho$) density for massive ($>$10$^{11}$ M$ _{\odot}$) galaxies (quiescent \& star-forming) in different environments since z$\sim$3. We start our analysis by studying the evolution of n \& $\rho$ for massive ($>$ 10$^{11}$ M$_{\odot}$) quiescent/star-forming system regardless of their environment. This allows us to compare n \& $\rho$ with previous studies which often do not consider the role of environment in their analysis. Later, we discuss the role of environment.\\
\begin{figure*}
  \centering
  \includegraphics[width=7.0in]{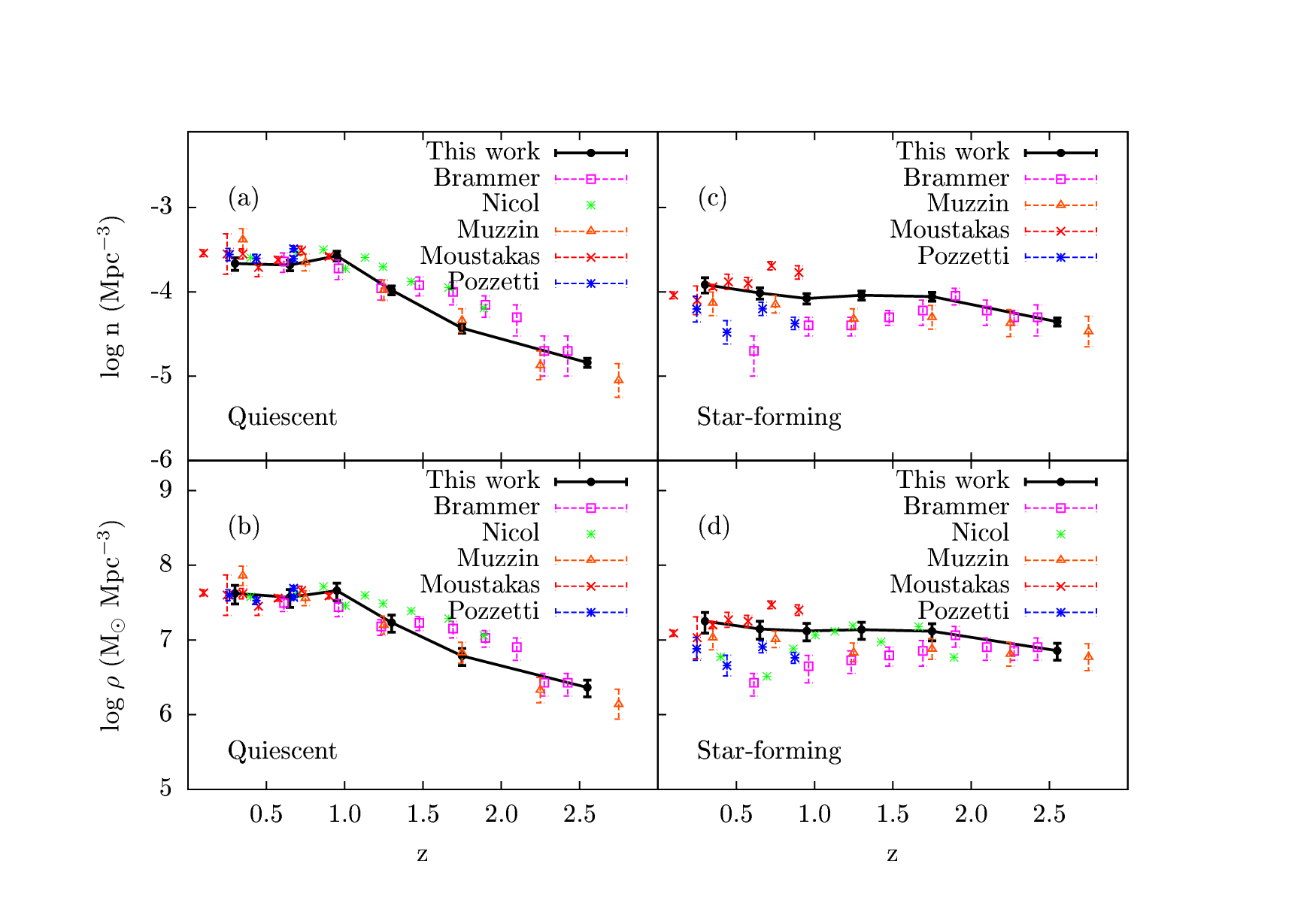}
\caption{Evolution of comoving number (n) and mass ($\rho$) density for massive ($>$10$^{11}$ M$ _{\odot}$) galaxies (quiescent \& star-forming) since z$\sim$3, along with comparison with some previous studies. Error bars for number densities incorporate both Poisson error and uncertainties due to the cosmic variance. Uncertainty in mass density is estimated using 10000 bootstrap resamples, added in quadrature to the observational uncertainties and those due to the cosmic variance. The comoving number and mass density of massive quiescent systems rapidly increase from z$\sim$3 to z$\sim$1. However, from z$\sim$1 to the present time, they remain almost unchanged within uncertainties. we find almost no evolution in the comoving number and mass density of massive star-forming galaxies since z$\sim$2. The number \& mass density of massive star-forming galaxies are slightly increased from z=2.0-3.1 to z=1.5-2.0 (a factor of $\sim$2) and remain almost unaltered since then.}
\label{fig:nmass-density}
\end{figure*}
Figure \ref{fig:nmass-density} shows the evolution of comoving number and mass density of massive ($> 10^{11} M_{\odot}$) quiescent and star-forming galaxies since z$\sim$3, along with comparison with some previous studies. Some of these studies use Kroupa IMF to estimate stellar masses. For those, we modify the mass densities based on the Chabrier IMF (M$_{Chab}$ $\sim$ 0.89 M$_{Kroupa}$). The redshift of our data points is selected as the center of redshift bins introduced for mass complete samples (Table 5). Error bars for number densities incorporate both Poisson error and uncertainties due to the cosmic variance. Uncertainty in mass density is estimated using 10000 bootstrap resamples, added in quadrature to the observational uncertainties and those due to the cosmic variance. For observational uncertainties, we use the median of the stellar mass uncertainties of the galaxies in each subsample. We use the cosmic variance calculator of \cite{Moster11} to estimate the cosmic variance uncertainties. For 11$<$log(M/M$_{\odot}$)$<$11.5 galaxies, the fractional uncertainties due to the cosmic variance change between $\sim$15-10\% at $\sim$z=0.1-3 in the COSMOS.\\
\begin{table*}
\begin{center}
{
\caption{Table 7: Comoving number density, n (in units of 10$^{-4}$ Mpc$^{-3}$) \& mass density, $\rho$ (in units of 10$^{7}$ M$_{\odot}$ Mpc$^{-3}$) of massive ($> 10^{11}$ M$_{\odot}$) quiescent \& star-forming galaxies in different redshift bins.} 
\begin{tabular}{lcccccc}
\hline
Property & \multicolumn{6}{c}{redshift} \\
\cline{2-7} 
         & 0.1$\leq$z$<$0.5 & 0.5$\leq$z$<$0.8 & 0.8$\leq$z$<$1.1 & 1.1$\leq$z$<$1.5 & 1.5$\leq$z$<$2.0 & 2.0$\leq$z$<$3.1 \\
\hline
n$_{Q}$ & 2.17$\pm$0.37 & 2.08$\pm$0.30 & 2.68$\pm$0.35 & 1.05$\pm$0.13 & 0.37$\pm$0.05 & 0.15$\pm$0.02 \\
n$_{SF}$ & 1.22$\pm$0.25 & 0.97$\pm$0.15 & 0.84$\pm$0.12 & 0.91$\pm$0.11 & 0.88$\pm$0.11 & 0.44$\pm$0.05 \\
$\rho_{Q}$ & 4.20$\pm$1.17 & 3.73$\pm$1.00 & 4.57$\pm$1.20 & 1.71$\pm$0.44 & 0.61$\pm$0.16 & 0.23$\pm$0.06 \\
$\rho_{SF}$ & 1.78$\pm$0.55 & 1.40$\pm$0.37 & 1.32$\pm$0.35 & 1.37$\pm$0.35 & 1.31$\pm$0.34 & 0.72$\pm$0.18 \\
\hline
\end{tabular}
\label{table:nmdensity}
}
\end{center}
\end{table*}

\subsubsection{Massive Quiescent Galaxies}

Figure \ref{fig:nmass-density} (a) and (b) show a rapid increase in the comoving number and mass density of massive quiescent systems from z$\sim$3 to z$\sim$1. However, we observe a change of pattern at z$\sim$1. From z$\sim$1 to the present time, the number \& mass density of massive quiescent galaxies remain almost unchanged within uncertainties. From z=2-3.1 to z=1.5-2.0, the number and mass density of massive quiescent galaxies are increased by a factor of $\sim$2.5. From z=1.5-2.0 to z=0.8-1.1, the rise in the number and mass density of massive quiescent galaxies is steeper and they are increased by a factor of $\sim$6-8.\\
The lack of evolution (or insignificant evolution) since z$\sim$1 in the comoving number and mass density of massive quiescent systems is consistent with several studies. In terms of sample selection, the closest works to our study are those of \cite{Ilbert13} \& \cite{Muzzin13}. \cite{Ilbert13} did not find any significant evolution in the number and mass density of high-mass end quiescent galaxies at z$<$1, consistent with our results. As seen in Figure \ref{fig:nmass-density} (a) \& (b), \cite{Muzzin13} data points follow our observed trends for n \& $\rho$ relatively well, within uncertainties. As seen in Figure \ref{fig:nmass-density} (a) \& (b), for quiescent galaxies at z$<$1, our result is also in a good agreement with $>$10$^{11}$M$_{\odot}$ red-sequence galaxies in COMBO17+4 Survey from \cite{Nicol11}, massive U-V-J selected quiescent galaxies in NMBS survey from \cite{Brammer11}, log(M/M$_{\odot}$)=11-11.5 red galaxies in zCOSMOS from \cite{Pozzetti10}, quiescent systems in PRIMUS survey from \cite{Moustakas13} and $>$10$^{11}$M$_{\odot}$ rest-frame (U-V) selected red galaxies in VIPERS from \cite{Davidzon13}. Several luminosity function studies at z$<$1 such as \cite{Scarlata07} \& \cite{Cool08} have seen almost no evolution in the number density of very luminous (L$\sim$2.5-3 L$^{*}$) red or early-type galaxies since z$\sim$1, in agreement with our results.\\
The sharp rise in the comoving number and mass density of massive quiescent systems from z=3-1 is also in agreement with some previous studies. For example, \cite{Ilbert13} found that the comoving number and mass density of 10$^{11}$ M$_{\odot}$ quiescent galaxies increase by factors 25 \& 13, respectively, between 2.5$<$z$<$3 \& 0.8$<$z$<$1.1. This is similar to our result showing an increase of $\sim$15-20 in n \& $\rho$ from z$\sim$3 to z$\sim$1 for massive quiescent systems. Our results also agree qualitatively/quantitatively with \citep{Nicol11,Brammer11,Dominguez-Sanchez11,Muzzin13}.
\begin{figure*}
  \centering
  \includegraphics[width=7.0in]{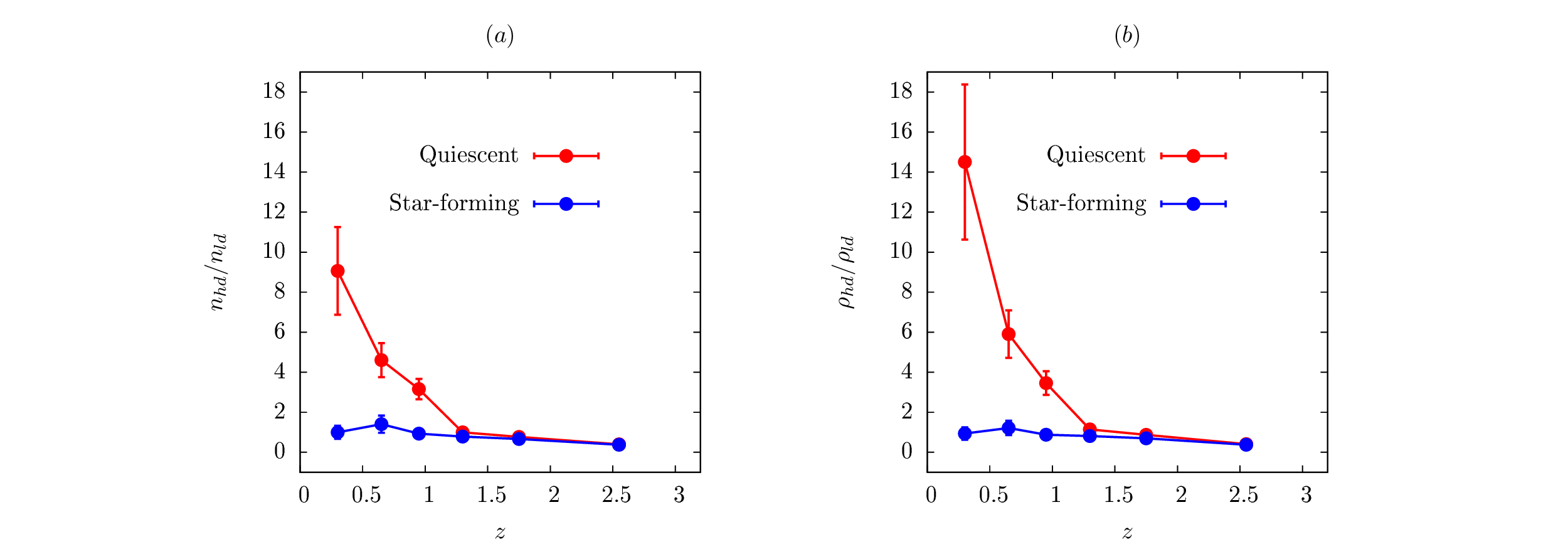}
\caption{(a) The ratio of the comoving number density in high-density environments to low-density environments (n$_{hd}$/n$_{ld}$) for massive ($>$ 10$^{11}$ M$_{\odot}$) quiescent and star-forming galaxies as a function of redshift. (b) The ratio of the stellar mass density in high-density environments to low-density environments ($\rho_{hd}$/$\rho_{ld}$) for massive ($>$ 10$^{11}$ M$_{\odot}$) quiescent and star-forming galaxies as a function of redshift. Both number and mass density ratios do not significantly change with redshift for massive star-forming galaxies. For massive quiescent systems, these ratios significantly change at lower redshifts, indicating the prevalence of massive quiescent galaxies in denser environments at lower redshifts.}
\label{fig:ratio}
\end{figure*}

\subsubsection{Massive Star-forming Galaxies}

The situation is different for massive ($>$10$^{11}$ M$_{\odot}$) star-forming systems. According to  Figure \ref{fig:nmass-density} (c) and (d), we find almost no evolution in the comoving number and mass density of star-forming galaxies since z$\sim$2. The number \& mass density of massive star-forming galaxies are slightly increased from z=2.0-3.1 to z=1.5-2.0 (a factor of $\sim$2) and remain almost unaltered since then. This slight (insignificant) increase is seen in \cite{Brammer11} from z$\sim$2.5-2.0, followed by decline since z$\sim$2. We argue that we do not find any sign of decline in the comoving number \& mass density of massive star-forming systems since z$\sim$2 in our study.\footnote[3]{\cite{Brammer11} state in their paper that the comoving number density evolution of $>$10$^{11}$ M$_{\odot}$ star-forming galaxies is nearly flat out to z=2.0}. Our result also agrees well with \citep{Pozzetti10,Muzzin13,Ilbert13,Davidzon13,Sobral14}. However, \cite{Moustakas13} found that the number (\& mass) density of massive star-forming galaxies are decreased slightly by a factor $\sim$2 since z$\sim$1 (Figure \ref{fig:nmass-density} (c) \& (d)). Within uncertainties, we do not find a significant disagreement between our result and that of \cite{Moustakas13}, although we detect no evolution in n \& $\rho$ for massive star-forming galaxies. We argue that even in the presence of a real evolution in number and mass density of massive star-forming systems since z$\sim$1, we would not be able to observe it due to the larger error bars in our study compared to \cite{Moustakas13}.\\
The comoving number and mass density values for massive ($>$ 10$^{11}$ M$_{\odot}$) quiescent/star-forming populations are given in table 7.

\subsubsection{Massive Quiescent/Star-forming Systems in Different Environments} \label{NMEQSF}

Now, we investigate the role of environment in shaping the number and mass density of massive quiescent and star-forming galaxies. In order to estimate comoving number and mass densities in different environments, one needs to determine what fraction of the density field is occupied by massive galaxies in different environments. This is not practically straightforward as the definition of environment and the selection of low- and high-density regions are arbitrary. We highlight that we only wish to \textit{relatively} compare the number and mass densities in different environments and not to determine the exact values in these regions. If there are any systematic errors in determining the number and mass densities in different environments, they would likely be cancelled out as we take the ratio of these in low- and high-density regions. We define the ratio of comoving number density of massive ($>$ 10$^{11}$ M$_{\odot}$) galaxies in high-density (n$_{hd}$) to low-density (n$_{ld}$) environments as:\\
\begin{equation}
\frac{n_{hd}}{n_{ld}}=\frac{\sum_{i} 1/V_{i}^{hd}}{\sum_{j} 1/V_{j}^{ld}}
\end{equation}
where V$_{i}^{hd}$ and V$_{j}^{ld}$ are the volumes associated with the $i$th and $j$th massive galaxy in high- and low-density environments, respectively. The volume assigned to each galaxy is estimated using its Voronoi area (A) and the radial comoving length ($\Delta$l) that corresponds to its z-slice; i.e. V$_{i}$=A$_{i}\Delta$l$_{i}$. Similarly, we define the ratio of mass density of massive ($>$ 10$^{11}$ M$_{\odot}$) galaxies in high-density ($\rho_{hd}$) to low-density ($\rho_{ld}$) environments as:\\
\begin{equation}
\frac{\rho_{hd}}{\rho_{ld}}=\frac{\sum_{i} M_{i}^{hd}/V_{i}^{hd}}{\sum_{j} M_{j}^{ld}/V_{j}^{ld}}
\end{equation}
where M$_{i}^{hd}$ and M$_{j}^{ld}$ are the stellar masses associated with the $i$th and $j$th massive galaxy in high- and low-density environments, respectively.\\
Figure \ref{fig:ratio} shows the redshift evolution of n$_{hd}$/n$_{ld}$ and $\rho_{hd}$/$\rho_{ld}$ for massive quiescent and star-forming galaxies. Uncertainties in number density ratios incorporate both Poisson error and the cosmic variance. The error bars for mass density ratios are estimated using 10000 bootstrap resamples, added in quadrature to the observational uncertainties in stellar mass and those due to the cosmic variance. For massive ($>$ 10$^{11}$ M$_{\odot}$) star-forming systems, the number and mass densities in different environments remain almost the same and we find almost no evolution in comoving number and mass density ratios in different environments with cosmic time. Massive star-forming galaxies populate dense and less-dense regions almost equally, regardless of their redshift. The situation is different for massive quiescent galaxies. At z$\gtrsim$1.3, there is no significant evolution in the comoving number and mass density ratios in different environments. At these redshifts, the number and mass density ratios for massive quiescent galaxies are almost equal to those of massive star-forming systems. However, at lower redshifts (z$\lesssim$1.3), we find a significant evolution in the comoving number and mass density ratios for massive quiescent systems in different environments. These ratios for massive quiescent galaxies monotonically increase with cosmic time at z$\lesssim$1.3. At z$\lesssim$0.5, the number, as well as mass density of massive quiescent galaxies are $\sim$1 dex higher in denser regions compared to less dense environments. Dense environments at low redshifts are populated by massive quiescent galaxies.\\
We stress that part of the evolution in the comoving number and mass density of massive quiescent systems in different environments might be due to the growth of the large scale structure with cosmic time. As time progresses, less dense environments eventually coalesce to assemble more massive, denser regions. We also mention that the lack of evolution at higher redshifts in the number and mass density ratios for massive quiescent galaxies in different environments does not necessarily mean that the environment at higher redshifts is not able to quench the star formation activity in galaxies as effectively as it does at lower redshifts. The likelihood of finding massive halos (that are able to suppress the star formation activity) at higher redshifts is low, given the small size of the COSMOS field and the rarity of massive halos at higher redshifts. If we could numerously find such dense massive halos at higher redshifts in our survey, we might be able to see similar environmental trends that we observe in lower redshifts and the local universe.\\
It is challenging to make a direct, quantitative comparison between our results in this section and similar studies in the literature, due to the differences in sample selections, the definition of environment and the classification of low- and high-density regions. However, we \textit{qualitatively} compare our results in this section with some similar studies. Using zCOSMOS data at z$<$1, \cite {Bolzonella10} studied the galaxy stellar mass function (SMF) as a function of galaxy type and environment. They found that massive galaxies are preferentially located in high-density environments, characterised on average by a higher M$^{*}$ and that the spectral-type selected early-type galaxies dominate the high-mass (log(M/M$_{\odot}$)$\gtrsim$10.7) end of the SMF. In other words, massive,red galaxies are preferentially reside in high-density regions at z$\lesssim$1, in agreement with our results for massive, quiescent galaxies in dense environments. Recently, \cite {Mortlock15} studied the SMF of galaxies in UKIDSS/UDS and CANDELS fields and found that at higher redshifts (z$>$1), the SMF parameters for galaxies in low and high densities are almost the same. However, at z$<$1, they found evidence that: 1) the high-mass end of the galaxy SMF is more dominated by galaxies in dense environments. 2) The high-mass end of the S\'{e}rsic index $>$ 2.5 SMF is dominated by quiescent galaxies. 3) on average, $\phi^{*}$ and M$^{*}$ values of the SMF are larger for red galaxies in denser regions compared to those in less-dense environments (At z$\sim$0.5, $\phi^{*}$ is $\sim$1 dex larger for red galaxies in denser regions, similar to our result in this section). Combining all these together, we conclude that at lower redshifts, high-density environments cause the build-up of high-mass, quiescent galaxies. This is completely consistent with our results in this section regarding the higher number and mass densities in denser regions for massive, quiescent systems at lower redshifts and the lack of any significant differences at higher redshifts.\\
We stress that our results here rely on the density field estimation based on the flux-limited sample (K$_{s}<$24) introduced in Section \ref{data}. However, as we already discussed in Section \ref{mass-comp}, this magnitude cut results in a sample with varying stellar mass as a function of redshift. In other words, the ``typical" galaxies defining the environment change with redshift, with less massive galaxies defining the low-z and more massive systems defining the high-z environments. Since there is some degree of correlation between galaxy stellar mass and environment \citep{Kauffmann04,Baldry06}, our results in this section might be affected by selection of galaxies used to estimate the density field (environment). However, in the Appendix, we investigate the role of different samples used for density estimation, on the results and will show that the results do not change (even become stronger) and we recover the same trends discussed in this section.\\  
We conclude that the comoving number and mass density of massive, star-forming systems do not evolve much with redshift, regardless of their environment. This scenario is also true for massive, quiescent galaxies at higher redshifts (z$\gtrsim$1.3). However, at lower redshifts, the comoving number and mass density of massive, quiescent galaxies are greater in high-density environments compared to less-dense regions. This highlights the significant role of environment in quenching the star formation activity in galaxies at lower redshifts.

\section{Summary \& Conclusions} \label{concl}

In this work, we used a K$_{s}$-band selected sample of galaxies with accurate photometric redshifts in COSMOS at z=0.1-3.1 in order to estimate the density field. The density field was determined with the weighted versions of the adaptive kernel smoothing, 10$^{th}$ \& 5$^{th}$ NN, Voronoi tessellation and Delaunay triangulation methods. We evaluated the performance of each density estimator using extensive realistic and Monte-Carlo simulations. We later defined two environments and studied the effects of environment on mass complete sample of quiescent and star-forming galaxies out to z$\sim$3. The Rest-frame NUV-r$^{+}$-J color-color plots were used to separate the galaxy population into quiescent and star-forming systems. We investigated the redshift evolution of the rest-frame u$^{*}$- B$_{J}$ color for quiescent galaxies as a function of environment and stellar mass. We also studied the evolution of comoving number and mass density of massive quiescent and star-forming galaxies and their dependence on the environment of galaxies. Our main results are:\\

1- We find an overall good agreement between the density field estimated with weighted versions of adaptive kernel smoothing, nearest neighbor (NN), Voronoi tessellation and Delaunay triangulation methods over $\sim$2 orders of magnitude.\\

2- Extensive simulations show that the adaptive kernel smoothing and Voronoi tessellation outperform other methods in estimating the density field of galaxies. We recommend using these estimators as a more reliable and robust substitute for the widely-used NN or count in aperture methods.\\

3- At fixed stellar mass, the median rest-frame u$^{*}$-B$_{J}$ color of quiescent galaxies is independent of the environment they reside in. Quiescent galaxies become redder with cosmic time and their color evolution is independent of their environment. Since z$\sim$3, more massive quiescent galaxies (log(M/M$_{\odot}$)$\gtrsim$10) have become $\sim$0.2 mag redder in rest-frame u$^{*}$-B$_{J}$ whereas less massive quiescent systems have reddened by $\sim$0.1 mag since z$\sim$1.\\
 
4- On average, more massive quiescent galaxies are redder compared to less massive ones at any given redshift, regardless of their environment. The lack of a correlation between color and environment at fixed stellar mass for quiescent galaxies suggests that the relation between stellar mass and local density of galaxies is more fundamental than the color-density relation.\\

5- The average comoving number and mass density of massive (log(M/M$_{\odot}$)$>$11) star-forming galaxies have not evolved much since z$\sim$3. However, for massive quiescent galaxies, number and mass densities sharply rise from z$\sim$3 to z$\sim$1 and remain almost unchanged since then.\\

6- The evolution of comoving number and mass density of massive star-forming galaxies do not depend on their environment. They remain almost unchanged since z$\sim$3, regardless of their host environment. The situation is different for massive quiescent galaxies. The comoving number and mass density of massive quiescent galaxies do not change much with environment and redshift from z$\sim$3 to z$\sim$1.3, similar to those of star-forming galaxies. However, at lower redshifts (z$\lesssim$1.3), we find a significant evolution in the number and mass density of massive quiescent galaxies in denser environments compared to less-dense regions. Dense environments at lower redshifts are populated with massive quiescent galaxies. This signifies the role of environment in quenching the star formation activity in galaxies at lower redshifts.\\

This paper is the first one in a series and provided the required tools for the density-based environmental study of galaxies. In a following paper (in prep.), we will study the effects of local environment of galaxies on their SFR, sSFR, rest-frame color and the fraction of quiescent/star-forming systems as a function of redshift, stellar mass and galaxy type. We will also discuss the fractional role of stellar mass and environment in suppressing the star formation activity in galaxies.

\section*{acknowledgements}

We gratefully thank the anonymous referee for thoroughly reading the original manuscript and providing very useful comments that improved the quality of the work. B.D. gratefully acknowledges Lucia Pozzetti for providing the data for this study. D.S. acknowledges financial support from LKBF, the Netherlands Organisation for Scientific research (NWO) through a Veni fellowship, from FCT through an FCT Investigator Starting Grant, a Start-up Grant (IF/01154/2012/CP0189/CT0010), and the grant PEst-OE/FIS/UI2751/2014. This study has used the COSMOS data based on observations with the NASA/ESA Hubble Space Telescope, obtained at the Space Telescope Science Institute, which is operated by AURA Inc., under NASA contract NAS 5-26555, and the Spitzer Space Telescope, which is operated by the Jet Propulsion Laboratory, California Institute of Technology under NASA contract 1407; also based on data collected at the Subaru Telescope, which is operated by the National Astronomical Observatory of Japan; XMM-Newton, an ESA science mission with instruments and contributions directly funded by ESA Member States and NASA; the European Southern Observatory under Large Program 175.A-01279, Chile; Kitt Peak National Observatory, Cerro Tololo Inter-American Observatory, and the National Optical Astronomy Observatory, which are operated by the Association of Universities for Research in Astronomy, Inc. (AURA) under cooperative agreement with the National Science Foundation; the National Radio Astronomy Observatory, which is a facility of the National Science Foundation operated under cooperative agreement by Associated Universities, Inc.; and the Canada–France–Hawaii Telescope with MegaPrime/MegaCam operated as a joint project by the CFHT Corporation, CEA/DAPNIA, the NRC and CADC of Canada, the CNRS of France, TERAPIX, and the University of Hawaii.

\appendix

\section{Effect of Density Field Estimation based on Different Sample Selections}  
In Section \ref{data}, we used a flux-limited (K$_{s}<$24) sample of galaxies in order to estimate the density field. All the main results presented in Sections \ref{CMEZ} \& \ref{NMEQSF} were based on this flux-limited sample. Here, we try two other samples of galaxies to estimate the density field and investigate their effects on the results presented in Section \ref{NMEQSF}. Same as before, we use the Voronoi tessellation method as the density estimator and the definition of low- \& high-density environments given in Section \ref{CMEZ}. These new samples are defined as follows:
\begin{figure*}
  \centering
  \includegraphics[width=7.0in]{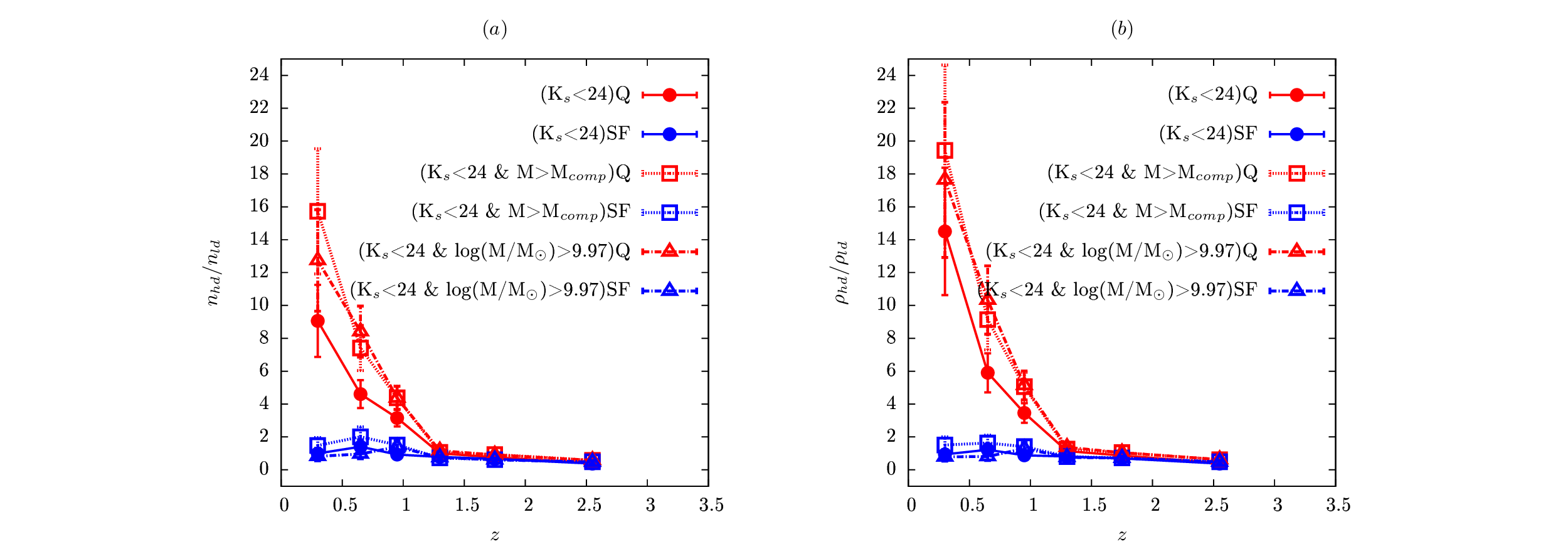}
\caption{Similar to Figure \ref{fig:ratio} but the results are shown for different samples used to estimate the density field of galaxies. Filled circles, empty squares and empty triangles show the trends based on a flux-limited sample (K$_{s}<$24, same results in Figure \ref{fig:ratio}), a sample of galaxies with (K$_{s}<$24 \& M$>$M$_{comp}$) and a sample of galaxies with (K$_{s}<$24 \& log(M/M$_{\odot}$)$>$9.97, similar to a volume-limited sample), respectively (see Appendix). We see the same trends obtained in Section \ref{NMEQSF}. Our results are not significantly influenced by the selection of the samples used to estimate the density field of galaxies.}
\label{fig:ratio-all}
\end{figure*}
\begin{enumerate}
\item In addition to conditions given in Section \ref{data}, we select all galaxies that are more massive than the stellar mass completeness limit of samples defined in Section \ref{mass-comp} and table 5 (K$_{s}<$24 \& M$>$M$_{comp}$).
\item In addition to conditions given in Section \ref{data}, we select all galaxies that are more massive than the stellar mass completeness limit of highest redshift sample defined in Section \ref{mass-comp} and table 5 (K$_{s}<$24 \& log(M/M$_{\odot}$)$>$9.97). This is similar to a volume-limited sample of galaxies.  
\end{enumerate}
Using these new samples, we re-estimate the density fields and re-investigate the results in Section \ref{NMEQSF}. Figure \ref{fig:ratio-all} is similar to Figure \ref{fig:ratio}, but it also shows the results based on the new samples used to estimate the density field. According to Figure \ref{fig:ratio-all}, we are able to retrieve the main trends obtained in Section \ref{NMEQSF} for quiescent (Q) and star-forming (SF) galaxies. However, we mention that for massive quiescent galaxies, the number and mass density trends become even more amplified at z$<$1 when we use the new samples for density estimation. This is expected since the new samples target more massive galaxies in order to estimate the density field, which tend to be more strongly clustered compared to less massive systems. We highlight that our results in Section \ref{NMEQSF} are not significantly affected by the selection of the samples we used for density field estimation.

\bibliographystyle{apj} 
\bibliography{densityref}

\end{document}